\documentclass[12pt]{article}
\usepackage[english]{babel}
\usepackage[latin1]{inputenc}

\usepackage[intlimits]{amsmath}
\usepackage{amssymb}

\usepackage{url}
\usepackage{graphicx}
\usepackage{slashed}
\usepackage{color}
\usepackage{cite}

\numberwithin{equation}{section}

\newcommand{\TeV}{\mbox{TeV}}

\begin{document}

\title{\vspace{-2cm} 
{\normalsize
\flushright CERN-PH-TH/2014-041 \\
\vspace{-.4cm}\flushright TUM-HEP 935/14\\
\vspace{-.4cm}\flushright DESY 14-029\\}
\vspace{0.6cm} 
\bf Majorana Dark Matter with a Coloured Mediator: Collider vs Direct and Indirect Searches\\[8mm]}

\author{Mathias Garny$^1$, Alejandro Ibarra$^2$, Sara Rydbeck$^3$, Stefan Vogl$^2$\\[2mm]
{\normalsize\it  $^1$ CERN Theory division,}\\[-0.05cm]
{\normalsize\it  CH-1211 Geneva 23, Switzerland}\\[1.5mm]
{\normalsize\it $^2$Physik-Department T30d, Technische Universit\"at M\"unchen,}\\[-0.05cm]
{\it\normalsize James-Franck-Stra\ss{}e, 85748 Garching, Germany}\\[1.5mm]
{\normalsize\it  $^3$ Deutsches Elektronen-Synchrotron DESY,}\\[-0.05cm]
{\normalsize\it  Notkestra\ss{}e 85, 22603 Hamburg, Germany}
}

\maketitle

\begin{abstract}
\noindent
We investigate the signatures at the Large Hadron Collider of a  minimal model where the dark matter particle is a Majorana fermion that couples to the Standard Model via one or several coloured mediators. We emphasize the importance of the production channel of coloured scalars  through the exchange of a dark matter particle in the t-channel, and perform a dedicated analysis of searches for jets and missing energy for this model. We find that the collider constraints are highly competitive compared to direct detection, and can even be considerably stronger over a wide range of parameters. We also discuss the complementarity  with searches for spectral features at gamma-ray telescopes and comment on the possibility of several coloured mediators, which is further constrained by flavour observables.
\end{abstract}

\newpage

\section{Introduction}

Despite the mounting evidence for the existence of dark matter (DM) in galaxies, clusters of galaxies and the Universe at large scale, the nature and properties of the dark matter particle are still largely unconstrained by observations. In fact, viable dark matter models have been constructed with masses ranging between $\sim 1~\mu$eV and $\sim 10^{16}$ GeV, and interaction cross sections ranging between $\sim 10^{-35}$ pb and $\sim 1$ pb (for a review, see \cite{Bertone:2004pz}). In this vast parameter space of dark matter models, Weakly Interacting Massive Particles (WIMPs) still stand as one of the most promising dark matter candidates,  since for reasonable values of the model parameters, the freeze-out of dark matter WIMPs from the thermal plasma left a relic population with an abundance which   reproduces qualitatively well the measured value of the dark matter density $\Omega_{\rm DM} h^2=0.1199\pm 0.0027$ \cite{PlanckCollaboration2013}.

There are presently three different approaches pursued in order to detect the non-gravitational effects of WIMPs with ordinary matter: direct detection, indirect detection and collider experiments. This decade is being especially prolific in experimental results in the three search strategies. Indeed, various experiments currently in operation are setting strong limits on the WIMP parameter space and  ruling out regions where a dark matter signal could be expected, notably XENON100~\cite{Aprile:2012nq} and LUX~\cite{Akerib:2013tjd} in direct searches, Fermi-LAT~\cite{Fermi-LAT:2013uma}, AMS-02~\cite{Aguilar:2013qda}, H.E.S.S.~\cite{Abramowski:2013ax}, MAGIC~\cite{Aleksic:2013xea}, IceCube~\cite{Aartsen:2012kia} in indirect searches and the LHC in collider searches (see e.g. \cite{CMS:zxa,TheATLAScollaboration:2013fha,Chatrchyan:2012me,ATLAS:2012zim}). Moreover, in the near future the $14$\,TeV run of LHC, the XENON1T~\cite{Aprile:2012zx} and LZ~\cite{Malling:2011va} experiments, and the Cerenkov Telescope Array~\cite{Bernlohr:2012we} will significantly improve the reach of collider, direct and indirect dark matter searches, respectively. 

These three different approaches constrain the parameter space of dark matter models in a complementary way, however, the synergy of the various search strategies is very model dependent. In this paper we focus on a simple scenario where the dark matter particle is a Majorana fermion that couples to light quarks and a coloured scalar via a Yukawa coupling. This scenario, despite its simplicity, offers a very rich phenomenology in direct detection \cite{Hisano:2011um,Garny:2012eb,Hamaguchi:2014pja},  indirect detection \cite{Garny:2011ii,Asano:2011ik,Bringmann:2012vr,Garny:2013ama,Ibarra:2013eba,Ibarra:2014vya} and collider experiments \cite{Chang:2013oia,An:2013xka,Bai:2013iqa,DiFranzo:2013vra}. In particular, when the mediator mass is comparable to  the dark matter mass, this model predicts a sharp and relatively intense gamma-ray spectral feature which, if observed, would constitute an unambiguous signal for dark matter annihilations \cite{Bergstrom:1989jr,Bringmann:2007nk,Bringmann:2011ye}. Additionally, the collider phenomenology is distinct from the widely-used effective operator approach (see e.g. \cite{Rajaraman:2011wf,Fox:2011pm,Busoni:2013lha}), because the mediator can be directly produced in proton proton collisions. Similar models, but with leptonic mediators, were studied in \cite{Ciafaloni:2011sa,Bell:2011if,Garny:2011cj,Bringmann:2012vr,Garny:2013ama,Kopp:2014tsa,Chang:2014tea,Bai:2014osa}. 

In this paper we revisit the collider limits in this scenario. Most analyses include only the production of coloured scalars via strong interactions, nevertheless, in this scenario the Yukawa coupling can be sizeable and the production of coloured scalars via the exchange of a dark matter particle in the t-channel can become important or even dominant. This possibility has been discussed in \cite{Chang:2013oia,An:2013xka,Bai:2013iqa,DiFranzo:2013vra}.  Here we go beyond these analyses by performing a dedicated re-interpretation of collider searches which includes also jet matching, that is important when considering the quasi-degenerate mass spectrum. A similar analysis for the case of Dirac dark matter has been recently presented in \cite{Papucci:2014iwa}. We analyse the limits on the Yukawa coupling from  the ATLAS search for jets and missing transverse energy \cite{TheATLAScollaboration:2013fha} and investigate the complementarity of the collider  limits with those from direct and indirect dark matter searches.  Furthermore we discuss  various sources of experimental and theoretical uncertainties of collider limits and assess their impact on the exclusion power. Finally, we consider an extension of the model by two coloured scalars coupling to the up-type quarks and we study the impact of extending the scalar sector on the dark matter searches in view of the stringent limits from flavour violation.

The paper is organized as follows. In section \ref{sec:model}, we introduce the simplified model and discuss its properties with respect to indirect, direct and collider searches. Section \ref{sec:LHC} explains some details of our collider analysis. Our results are discussed and compared to direct and indirect detection constraints in section \ref{sec:results}, and we conclude in section \ref{sec:conclusions}. The Appendix contains a brief discussion of flavour constraints.

\section{Particle physics model and observables}\label{sec:model}

 We assume the dark matter particle $\chi$ to be a Majorana fermion which couples to the light quarks via a Yukawa interaction with  
coloured scalars $\eta_i$. The Lagrangian of the model can be written as 
\begin{equation}\label{eq:L}
\mathcal{L}= \mathcal{L}_{\rm SM} + \mathcal{L}_{\chi} + \mathcal{L}_{\eta} + \mathcal{L}_{\rm int},
\end{equation}
 where $\mathcal{L}_{\rm SM}$ denotes the Standard Model (SM) Lagrangian while  $\mathcal{L}_{\chi}$ and $\mathcal{L}_{\eta}$ are given by
\begin{align}
  \begin{split}
    {\cal L}_\chi&=\frac12 \bar \chi^c i\slashed {\partial} \chi
    -\frac{1}{2}m_{\chi} \bar \chi^c\chi\;, \; \text{and}\\ 
    {\cal L}_\eta&=(D_\mu
    \eta_i)^\dagger  (D^\mu \eta_i)-m_{\eta_i}^2 \eta_i^\dagger\eta_i \;,
  \end{split}
\end{align}
where $D_\mu$ denotes the 
covariant derivative. On the other hand, $\mathcal{L}_{\rm int}$ contains the interactions between the SM quarks and the dark sector,
 \begin{align}
 {\cal L}_{\rm int} &=  - f_{ij} \bar q_{R i} \chi \eta_{j}+{\rm h.c.} \;,
\label{eq:singlet-eR}
\end{align}
where $f_{ij}$ is a Yukawa coupling matrix, $q_{R i} $ denote the right-handed quark fields and summation over flavours $i$, $j$ is implied. This Lagrangian generically leads to too large flavour changing neutral currents, hence some requirements must be imposed on the Yukawa couplings to fulfil the stringent constraints from flavour observables. In the following we consider two scenarios: 
\begin{enumerate}
\item We consider a single scalar $\eta$ that couples exclusively to the right-handed up quarks, with coupling strength $f$. This scenario corresponds to an alignment type set-up of the squark sector in the MSSM  and can be realized by appropriate flavour symmetries at a high scale~\cite{Nir:1993mx}.
\item We consider a pair of mass degenerate scalars $\eta_u$ and $\eta_c$ which couple to right-handed up and charm quarks with a universal coupling $f_{ij}=\delta_{ij} f$. Such a scenario is motivated by the paradigm of minimal flavour violation~\cite{D'Ambrosio:2002ex} which requires flavour universality among quarks with the same gauge quantum numbers while allowing a separation of particles belonging to different multiplets.  
\end{enumerate} 

We show explicitly in Appendix \ref{App:flavour} that within these two scenarios  the constraints from flavour observables are easily satisfied. One may also consider a coupling to down-type quarks, which is completely analogous and qualitatively very similar. In the following we  concentrate on the above two scenarios for definiteness. The model is thus completely described by the two masses $m_{\chi}$ and $m_\eta$ of the dark matter and the mediator(s), respectively, and by the Yukawa coupling $f$. With this framework it is possible to calculate various dark matter observables, e.g. the relic density, the annihilation cross section, the  dark matter\--nucleon scattering cross section or event rates at the LHC, and compare their relative exclusion power.

An interesting particularity of the model analyzed here is that the strongest experimental constraints  can not be derived from a small set of effective operators, but require to consider higher order effects. Concretely, for indirect detection the two-to-three and loop-induced annihilation channels play an important role, firstly because the leading order two-to-two channel is helicity  and velocity suppressed, and second because the hard gamma-ray spectrum from $\chi\chi\to q\bar q\gamma$ and the loop induced processes $\chi\chi\to \gamma\gamma,\gamma Z$ generate a very distinct spectral signature  \cite{Bergstrom:1989jr,Bringmann:2007nk,Bringmann:2012vr,Garny:2013ama}. For direct detection, the lowest order operators mediating spin-independent interactions are suppressed for Majorana dark matter with chiral interactions, such that higher order contributions and the spin-dependent scattering have to be also considered \cite{Hisano:2011um,Garny:2012eb}. Lastly, the production at the LHC is governed not only by the strong processes, but also by processes mediated by the Yukawa interaction with the dark matter particle~\cite{Chang:2013oia,An:2013xka,Bai:2013iqa,DiFranzo:2013vra}.

In the following, we summarize the relevant features of the model concerning the relic density, as well as the direct and indirect detection,\footnote{A detailed discussion including expressions for the spin dependent (SD) and spin  independent (SI) scattering cross section can be found in \cite{Garny:2012eb,Garny:2013ama}.} and then discuss  in detail the signatures at the LHC. 
 
\subsection{Thermal relic density}

Probably the most compelling argument for WIMP dark matter is that this class of particles 
is produced quite naturally in the early Universe and can generate, after thermal freeze-out, the 
correct relic density   $\Omega_{\rm DM} h^2=0.1199\pm 0.0027$ \cite{PlanckCollaboration2013} as measured by the Planck satellite.
The Lagrangian (\ref{eq:L}) allows for tree level annihilations $\chi \chi \rightarrow q \bar{q}$ and in most of the parameter space the relic abundance is set by this process.
However, for $m_{\eta}/m_{\chi} \lesssim 1.2$ the scalar $\eta$ does not freeze-out before the dark matter particle $\chi$
and modifies the relic density \cite{Griest:1990kh}.  This process, which is known as coannihilation, can be approximately taken into account by first defining an effective cross section
\begin{equation}
\sigma v_{\rm eff}=\sigma v(\chi \chi) + \sigma v (\chi \eta) e^{-\frac{m_\eta - m_{\chi}  }{T_{\rm fo}} }+ \sigma v (\eta \eta) e^{-\frac{2 (m_\eta - m_{\chi})}{T_{\rm fo}}}\;,
\end{equation}  
where $T_{\rm fo}$ corresponds to the freeze-out temperature while $\sigma v (\chi \eta) $ and $\sigma v (\eta \eta)$ correspond to the thermally averaged annihilation  cross-section of a $\chi \eta$ or an $\eta \eta$ pair respectively, and then replacing the thermally averaged cross section by this effective cross section in the well-known solution to the Boltzmann equation neglecting coannihilations. 
In our analysis, we use  micrOMEGAs2.4 \cite{Belanger:2010gh} to calculate the relic density in a full numerical approach (see also \cite{deSimone:2014pda} for a recent discussion of Sommerfeld enhancement in a similar context).

\subsection{Indirect detection}

\begin{figure}
\hspace*{-0.3cm}
\begin{tabular}{cc}
 \includegraphics[width=0.47\textwidth]{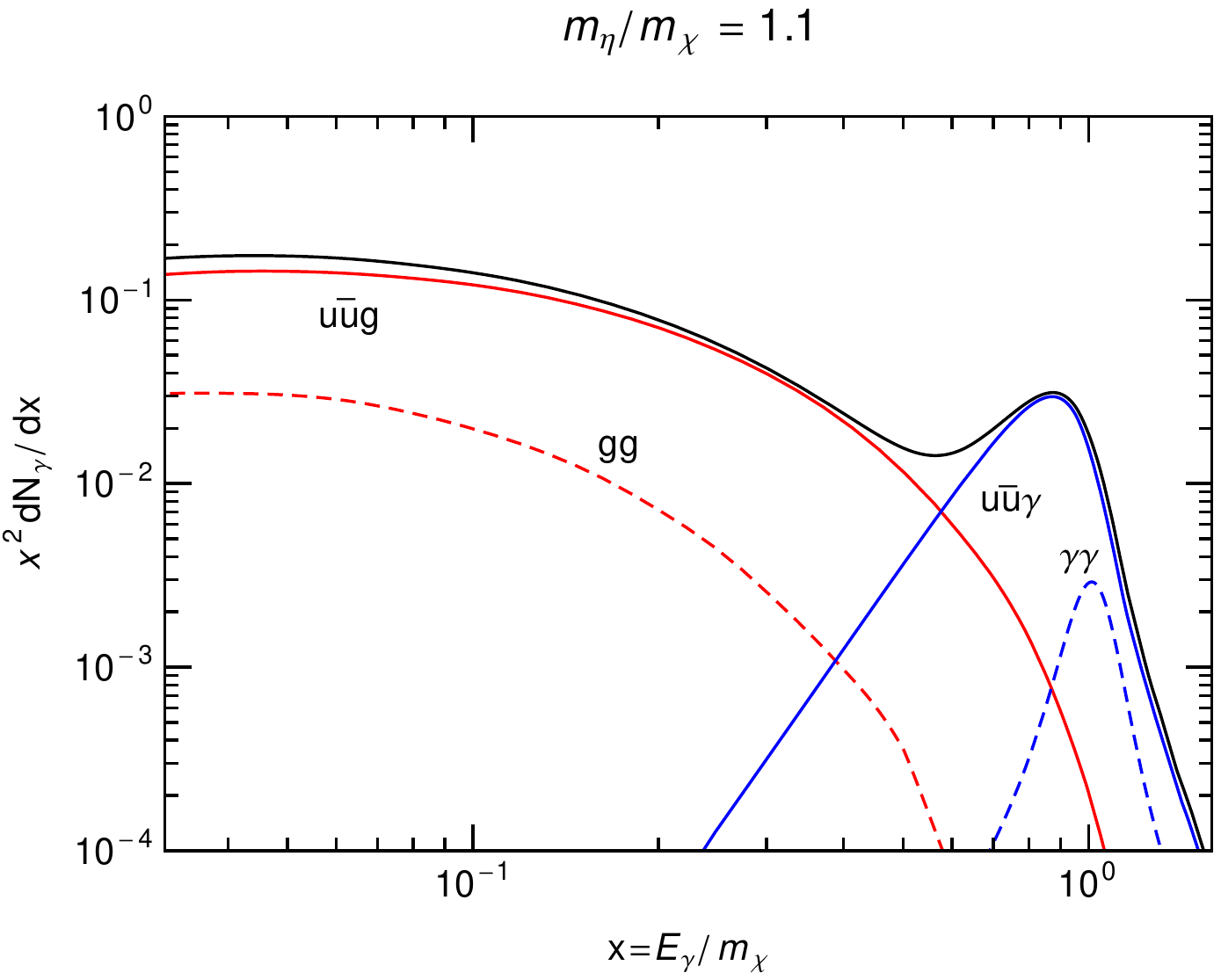} &
 \includegraphics[width=0.47\textwidth]{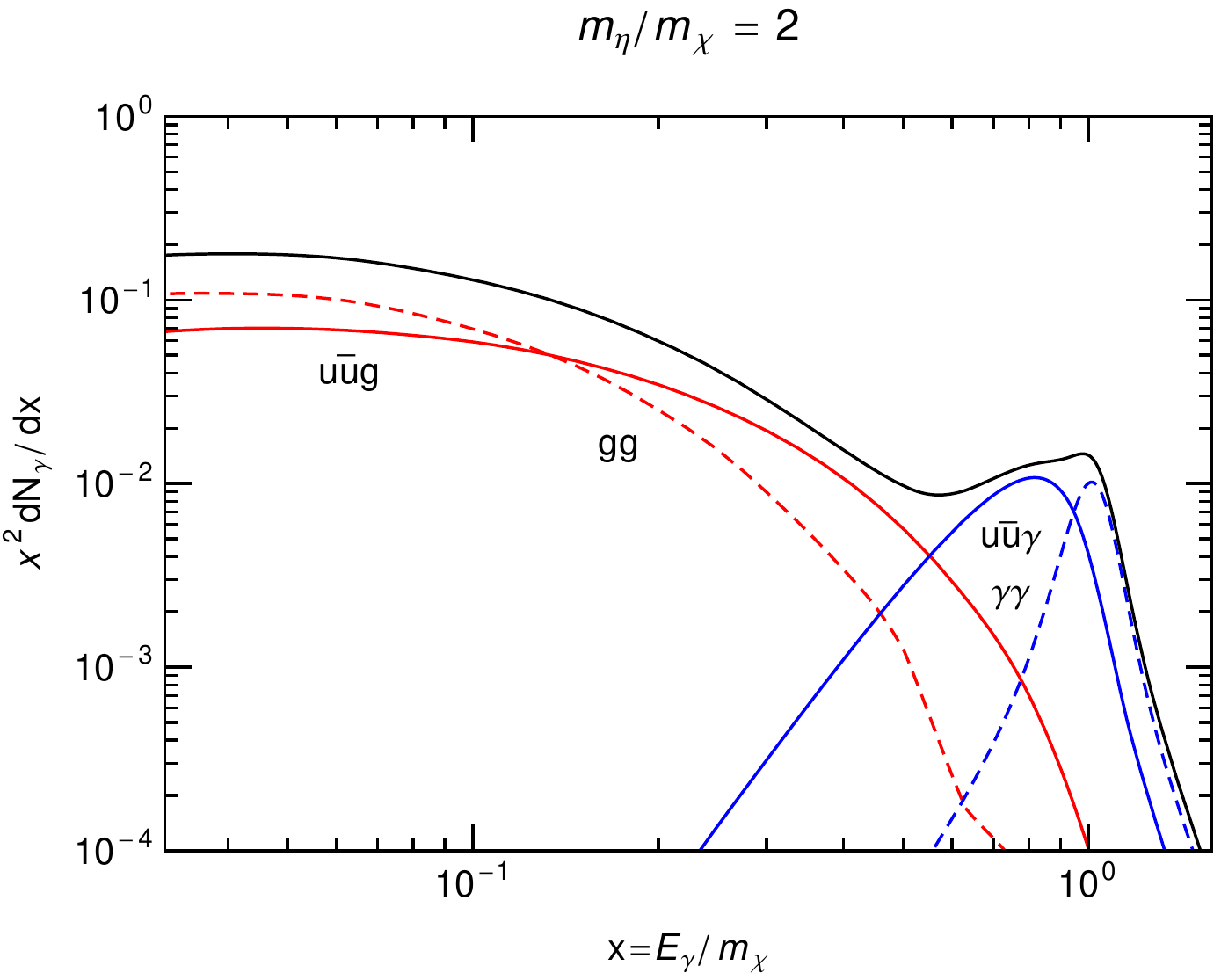}
\end{tabular}
 \caption{\label{fig:spectrum} Energy spectrum of gamma rays produced in the annihilation channels $u\bar u\gamma$, $\gamma\gamma$ as well as $u\bar ug$ and $gg$ convolved with the Fermi-LAT energy resolution, for $m_{\chi}=100\,$GeV, $m_\eta=1.1m_{\chi}$ (left panel) and $m_\eta=2m_{\chi}$ (right panel). The black line indicates the total spectrum. Note that for $u\bar u\gamma$ only the primary spectrum is shown. The secondary gamma rays from this channel are negligible compared to those arising from $u\bar ug$.}
\end{figure}

The processes relevant for dark matter annihilations in the galaxy today are closely related to those that determined the thermal freeze-out in the early Universe. However, the freeze-out took place when the dark matter particles were still relativistic while annihilations today are a non-relativistic phenomenon. Expanding the tree level annihilation cross section for the two-to-two process for small values of the relative center of mass velocity, $v$,  one obtains \cite{EllisPhys.Lett.B444:367-3721998}
\begin{align}
  (\sigma v)_\text{2-body}= \frac{3 f^4 }{32\pi m_{\chi}^2}
  \frac{m_{q}^2}{m_{\chi}^2}\frac{1}{(1+\mu)^2}+{\cal O}(v^2)\;,
  \label{eqn:sv2s}
\end{align}
with $\mu = m_{\eta}^2/m_{\chi}^2 $ and $m_q$ the mass of the final state quark.  Therefore, the tree level annihilation cross section  for the two-to-two process is suppressed either by $m_{q}^2/m_{\chi}^2$, due to the helicity suppression\footnote{It is interesting to remark that the helicity suppression can be relaxed for a mediator with a small flavour off-diagonal coupling $f'$ to third-generation quarks, for which $\sigma v_{q\bar q'} \propto f^2{f'}^2$\,max\,$(m_q^2,m_{q'}^2)$, contrary to the naive expectation $\propto f^2{f'}^2m_qm_{q'}$. However, we will not pursue this possibility further here.}, or by the small velocity of the dark matter particles in the Galaxy today, $v\approx 10^{-3}$. As a result, higher order processes can potentially give the dominant contribution to the indirect detection signals.

In the model considered here two higher order processes become relevant. First, the loop induced annihilation into  two gauge bosons $\chi\chi\rightarrow \gamma\gamma,\,\gamma Z, \, ZZ,\, WW,\, gg$~\cite{Bergstrom1997,Bern:1997ng,Ullio:1997ke,Ibarra:2014vya},  and second, the  three-body annihilation $\chi \chi \rightarrow q \bar{q} V~ (q \bar{q} h)$, where a gauge boson $V$ (Higgs boson $h$) is emitted in association with the quarks \cite{Bergstrom:1989jr,Bringmann:2007nk,Bergstrom:2004cy,Bergstrom:2005ss,Flores:1989ru,Drees:1993bh, Bell:2011eu,Bell:2011if,Bell:2012dk,Ciafaloni:2011sa,Ciafaloni:2011gv,Ciafaloni:2012gs,DeSimone:2013gj, Garny:2011cj,Garny:2011ii,Asano:2011ik,Fukushima:2012sp,Luo:2013bua}.  These processes can dominate over the annihilation in $q \bar{q}$ pairs even though they are loop suppressed or three-body phase space suppressed, since they can generate velocity independent terms which do not suffer from helicity suppression. The analytical result for the cross sections into $\gamma\gamma$ and $\gamma Z$ is rather lengthy and we refer the reader to e.g. \cite{Bergstrom1997,Bern:1997ng,Ullio:1997ke}. On the other hand, the total annihilation cross section into two massless quarks and one photon is approximately given by \cite{Bell:2011if,Bringmann:2007nk}
\begin{align}
  (\sigma v)_\text{3-body}
  \simeq  &\frac{\alpha_\text{em} f^4 3 Q_q^2}{64\pi^2m_{\chi}^2}
   \left\{ 
  (\mu+1) \left[ \frac{\pi^2}{6}-\ln^2\left( \frac{\mu+1}{2\mu} \right)
  -2\text{Li}_2\left( \frac{\mu+1}{2\mu} \right)\right]  \right. \nonumber \\
   &+ \left. \frac{4\mu+3}{\mu+1}+\frac{4\mu^2-3\mu-1}{2\mu}\ln\left(
  \frac{\mu-1}{\mu+1} \right)
  \right\}\;,
  \label{eqn:sv3}
\end{align}
where $Q_q$ is the electric charge of the quark.

Due to the excellent energy resolution and the high statistics of present gamma-ray observatories, such as the Fermi-LAT or H.E.S.S., the most  notable indirect detection signature in this model is the hard gamma-ray spectral feature arising from the processes $\chi \chi \rightarrow \gamma \gamma$ and $\chi \chi \rightarrow q \bar{q} \gamma$ (the gamma-ray line from $\chi \chi \rightarrow \gamma Z$ is always subdominant). As $\chi \chi \rightarrow q \bar{q} \gamma$ scales like $1/\mu^4$ in the limit of large $\mu$, while $\chi \chi \rightarrow \gamma \gamma$ only scales as $1/\mu^2$, the former process is most relevant for moderate values of $\mu \lesssim 4$.  With current instruments the spectrum of internal bremsstrahlung is practically indistinguishable form a gamma-ray line, see Fig.\,~\ref{fig:spectrum}, hence it is necessary to derive limits on the combined spectrum based on data used in line searches. In the following we use the same procedure as in \cite{Garny:2013ama} and derive limits on the combined annihilation cross section into hard gamma-rays $\sigma v_{\rm combined}=2 \sigma v_{\gamma \gamma} + \sigma v_{q \bar{q} \gamma } $ employing data from the 
Fermi-LAT \cite{Bringmann:2012vr} and H.E.S.S. \cite{Abramowski:2013ax} observations of the galactic center region.

\subsection{Direct detection}\label{sec:DD}

The s-channel exchange of the scalar $\eta$ between the dark matter $\chi$ and the quark induces  spin dependent (SD) as well as spin independent (SI) scatterings off nucleons, while further contributions to the SI scattering arise from loop level interactions with both the quarks and the gluons \cite{Jungman}. On the one hand, the leading contribution to the SD interactions can be described by a dimension-six axial-vector dark matter quark interaction ~\cite{DreesNojiri}
\begin{eqnarray}
\mathcal{L}_{\rm eff}^{\rm SD} = d_q \bar{\chi} \gamma^{\mu} \gamma^5 \chi \bar{q} \gamma_{\mu} \gamma^5 q \;,
\end{eqnarray}     
where the dark matter coupling to the quarks  $d_q$ is a dimensional parameter which scales as
\begin{eqnarray}
d_{q} \propto \frac{f^2}{m_{\eta}^2- (m_{\chi} + m_q)^2}\;.
\end{eqnarray}
The case of SI scattering is, on the other hand, more involved  since the coefficient $f_q$ of the scalar term in the effective Lagrangian ${\cal L}_{\rm eff,scalar}^{ \rm SI} =  f_q \bar{\chi} \chi \bar{q} q $ vanishes at dimension-six for chiral interactions, while  the coefficient for the vector interactions $\bar{\chi} \gamma^{\mu} \chi \bar{q} \gamma_{\mu} q$ vanishes to all orders, due to $\chi$ being a  Majorana fermion. Consequently the leading contribution to the SI coupling between the dark matter and the nucleons is generated at higher order. Expanding the scalar exchange beyond dimension six, the first non vanishing contribution arises at dimension eight \cite{DreesNojiri}; the strength of this interaction is proportional to 
\begin{eqnarray}
g_{q} \propto \frac{f^2}{(m_{\eta}^2- (m_{\chi} + m_q)^2)^2}\;.
\end{eqnarray}      
A further, subdominant, contribution to the dark matter nucleus coupling is induced by the scattering of the dark matter off the gluon content of the nucleon via a scalar-quark loop, which generates a dimension-seven effective operator \cite{DreesNojiri}.

Due to the structure of $d_q$ and $g_q$, a small mass difference $\Delta m = m_{\eta} - m_{\chi}$ between the scalar particle $\eta$ and the dark matter particle $\chi$ can lead to a drastic enhancement of both the SD and SI  scattering cross sections \cite{Hisano:2011um,Garny:2012eb,Garny:2012it} (see \cite{Gondolo:2013wwa} for a recent discussion of coupling to b-quarks).  Since the SI scattering is generated by a  higher order operator than the SD scattering cross section, the SD dark matter-proton cross section $\sigma^{\rm SD}_p$ can exceed the SI dark matter-proton cross section $\sigma^{\rm SI}_p$ by a factor $\sim 10^2-10^8 $ at $m_{\eta}/m_{\chi}=1.1$. However, since direct detection experiments are less sensitive to the SD scattering, the strongest constraints on the model parameters do not necessarily arise from the experimental limits on the SD interactions. 

In our analysis we consider constraints from two different experiments, LUX \cite{Akerib:2013tjd} and XENON100 \cite{Aprile:2012nq}. In the case of LUX we compare our theoretical prediction for the SI dark matter proton scattering cross section $\sigma^{\rm SI}_p$ directly with the limits published by the collaboration, while for XENON100 we use the same procedure as in \cite{Garny:2012eb}  and derive the limit from the total scattering rate,  {\it i.e} including both SD and SI contributions.\footnote{The LUX collaboration has not published limits on the total scattering rate, but derived their SI limits employing a full likelihood analysis of the recoil events, hence the approach in \cite{Garny:2012eb} cannot be applied to this case. }
Note that SIMPLE and COUPP \cite{SIMPLE11,COUPP12} yield comparable constraints for cases where the SD part dominates, since the scattering cross section off neutrons and protons is comparable for the model considered here.

\subsection{Production at LHC}\label{sec:prodLHC}

Searching for dark matter at the LHC is inherently difficult as the production of particles which do not trigger a signal at the detectors can only be  investigated through the observation of large amounts of missing transverse energy (MET). In the recent past, mono-jet events with large MET have been a popular method to derive limits on the dark matter interactions with the SM particles \cite{Fox:2011pm,Rajaraman:2011wf}. However in models such as the one considered here it is possible to probe the dark sector more efficiently  by searching for the production of the mediator $\eta$ instead of the dark matter particle $\chi$. As $\eta$ is unstable and decays into a $\chi q$ pair, the expected signature of $\eta \bar{\eta} \; (\eta \eta)$ production are $n$-jet events, $n \geq 2$, with large missing energy, which allows for a better subtraction of the SM backgrounds than mono-jet events and consequently possess a higher sensitivity to new physics effects. Furthermore, since the mediator $\eta$ is coloured, it can be copiously produced in proton-proton collisions through strong interactions.

\begin{figure}
\begin{center}
 \includegraphics[width=0.95\textwidth]{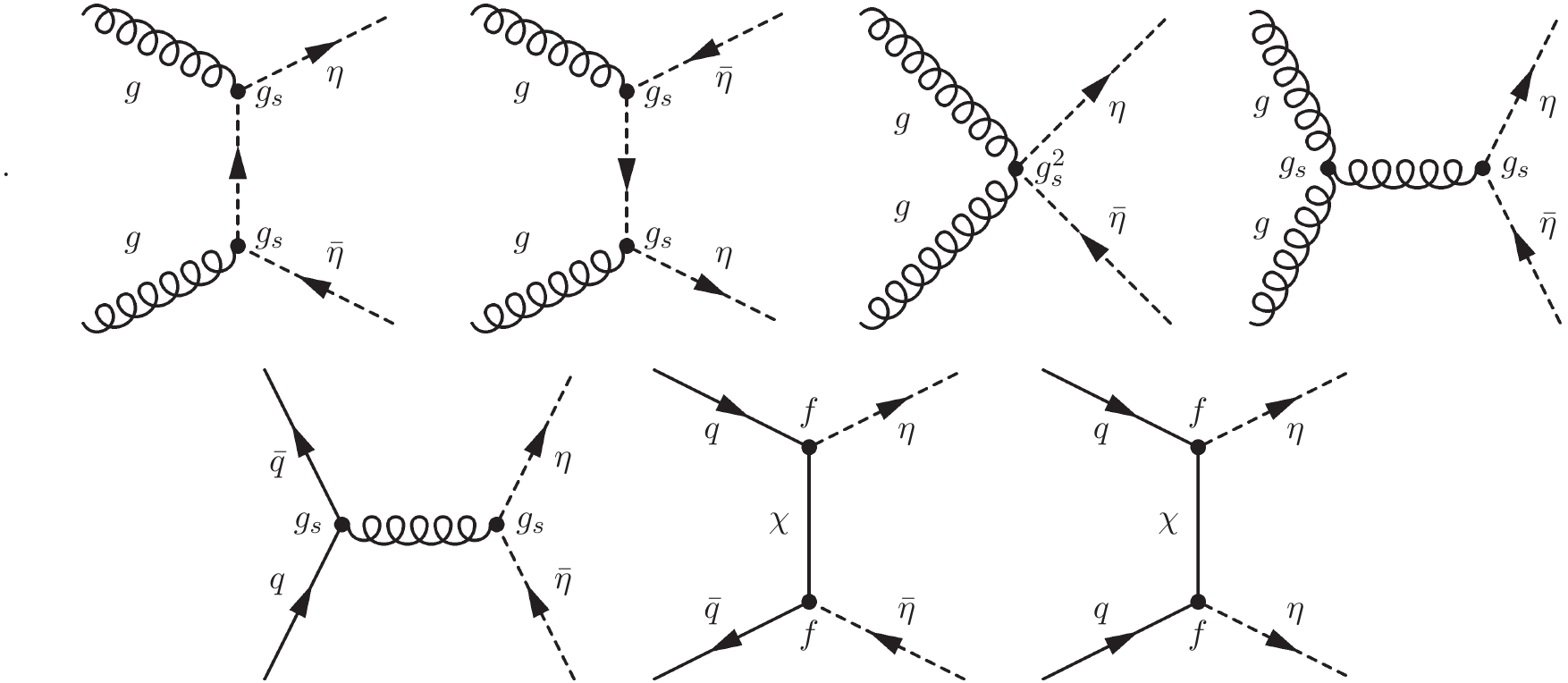}
\end{center}
 \caption{\label{fig:LHC} Feynman diagrams contributing to the production of coloured scalar mediators at a hadron collider.}
\end{figure}

\begin{figure}
\hspace*{-0.3cm}
\begin{tabular}{cc}
  \includegraphics[width=0.47\textwidth]{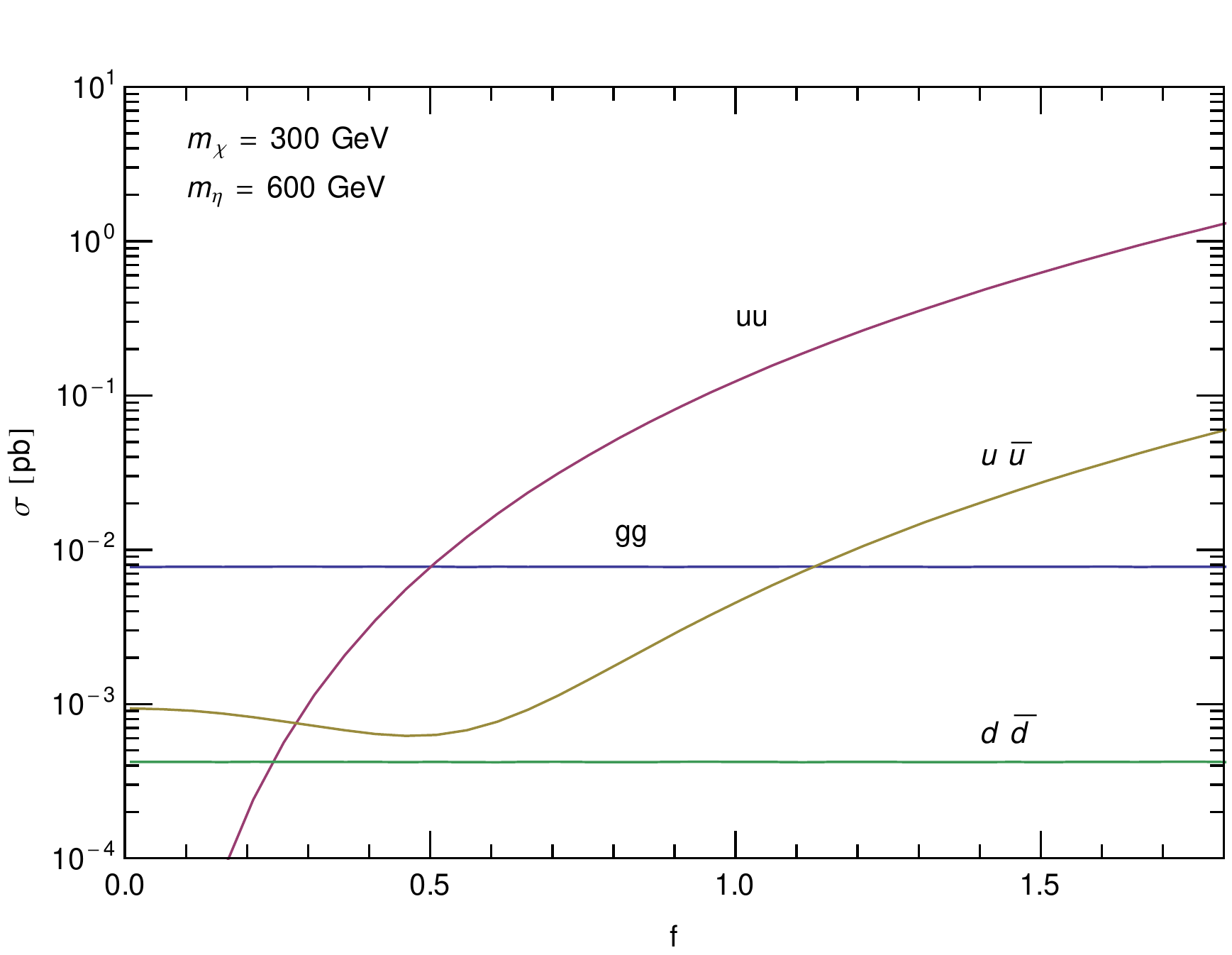} &
  \includegraphics[width=0.47\textwidth]{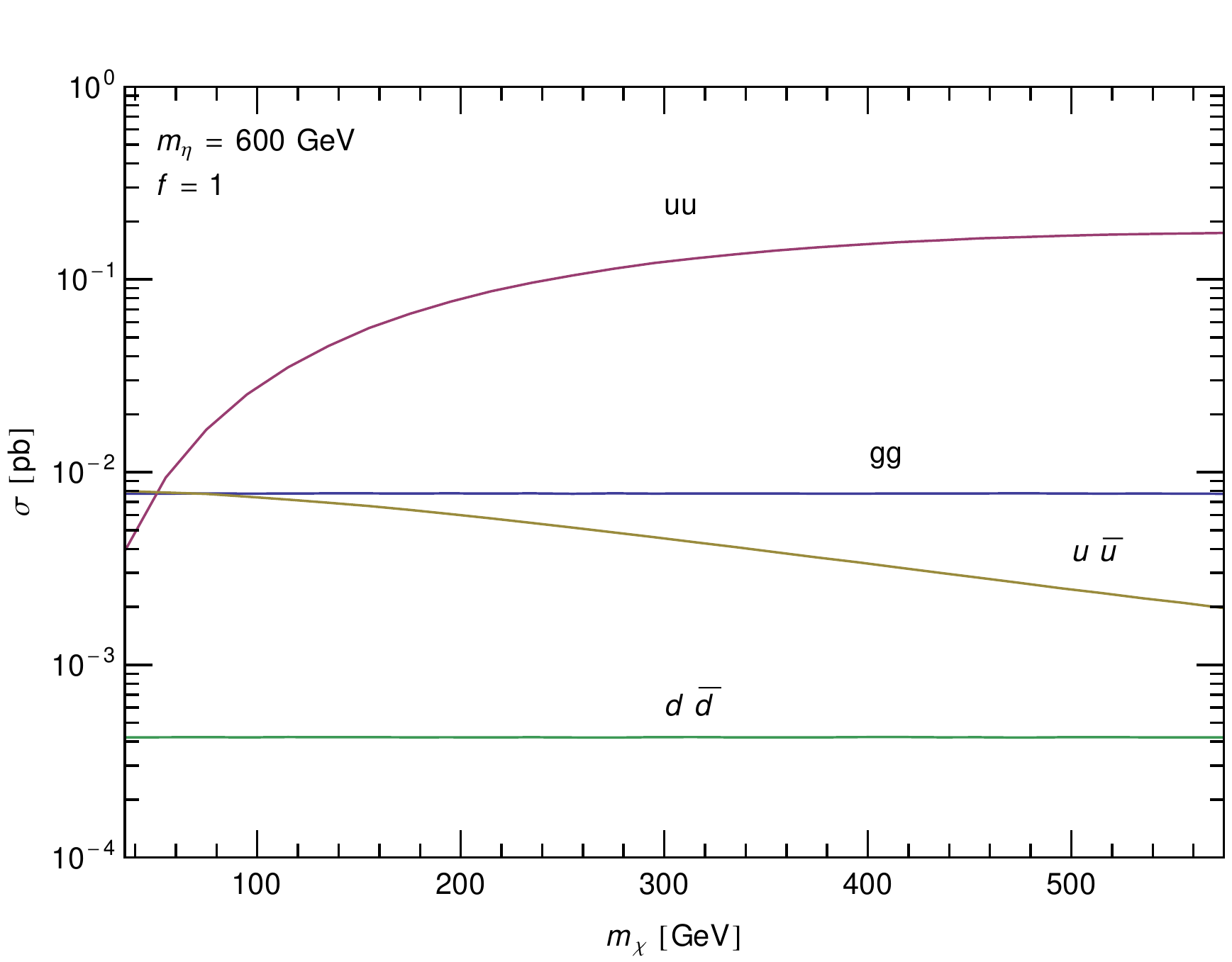}
\end{tabular}
  \caption{\label{fig:prod}
Contributions to the production cross section of the mediator $\eta$ in proton-proton collisions with center of mass energy of $8$\,TeV as a function of the  coupling $f$ for $m_{\chi}=300$\,GeV and $m_\eta=600$\,GeV (left panel) and as a function of the dark matter mass for fixed coupling $f=1$ and $m_\eta=600$\,GeV (right panel).}
\end{figure}

In principle there are three main production modes of $\eta$ which need to be considered in a collider analysis. The first process is the production of $\eta \bar{\eta}$ pairs by the strong gauge interactions from $g g $ or $q \bar{q}$  initial states, see Fig. \ref{fig:LHC} for the Feynman diagrams contributing to this process. The strong production cross section is set exclusively by the mass of $\eta$ and depends neither on $m_{\chi}$ nor on the Yukawa coupling $f$. As can be seen in Fig.~\ref{fig:prod}, due to the large gluon luminosity, the largest cross section arises from the $g g$ initial state with  $u \bar{u}$ and $d \bar{d}$ contributing at the $\mathcal{O}(1-10\%)$ level, while other quark flavours can be neglected.  A second contribution is the production of $\eta \bar{\eta} $ from the $u \bar{u}$ initial  state with a dark matter particle $\chi$ in the t-channel. This process is not independent but can interfere with the QCD contribution, see Fig.~\ref{fig:LHC}.  For small values of $f$ the cross section is largely dominated by the QCD processes, however for a moderate Yukawa coupling, $f \approx 0.5$, these two contributions are of similar strength and interfere destructively, thus leading to a slight decrease of the cross section (see Fig.~\ref{fig:prod}, left panel), while  for larger vales of $f$ the t-channel exchange begins to dominate and the cross section increases with $f^4$. The third contribution, $\eta \eta$ pair production, can not be induced by gauge interactions and is entirely due to the exchange of a dark matter particle in the t-channel. This process, which is similar to squark pair production from gaugino exchange in  the MSSM, requires a chirality flip of the t-channel fermion and is thus proportional to the squared mass of the dark matter particle $m_{\chi}^2$. Therefore the cross section $\sigma(\eta \eta)$ decreases with the DM mass and disappears  in the limit of a vanishing Majorana mass, whereas $\sigma(\eta \bar{\eta})$ increases slightly with lower $m_{\chi}$ since the t-channel exchange gets less suppressed by the mass in the propagator,  as apparent from Fig.~\ref{fig:prod}, right panel. Since  the parton distribution function  for the up quark  is significantly larger than for the anti-up quark, the cross section  $\sigma(\eta \eta)$ receives a considerable enhancement relative to  $\sigma(\eta \bar{\eta})$ and dominates the total cross section in large regions of the parameter space (see also \cite{Mahbubani:2012qq} for a similar effect related to gluino exchange).

Leading order calculations at the LHC are subject to large corrections from next to leading order (NLO) effects. While a full computation of the NLO corrected cross section $\sigma^{\rm NLO}$ is beyond the scope of this work, one can estimate its value using the well-known results from supersymmetric scenarios. Let us consider first the corrections to the partonic processes which do not receive a contribution from fermion t-channel exchange, namely $gg\rightarrow \eta \bar{\eta}$ and  $dd\rightarrow \eta \bar{\eta}$.  This case is completely analogous to the squark production in simplified supersymmetric models  where all SUSY particles,  except for the light squarks, are assumed to be decoupled.  For this case, the cross section is available at NLO and taking the resummation of next to leading logarithms (NLL) into account, $\sigma^{\rm NLO+NLL}_{\rm QCD}$ \cite{Kramer:2012bx}. In the following we use the value from \cite{LHCxsec}. On the other hand, the NLO correction to the channels $u u\rightarrow \eta\eta$ and $u \bar u\rightarrow \eta \bar \eta$,  where in addition to the strong production there is a contribution from the exchange of a dark matter particle in the t-channel, can be estimated from the results reported in \cite{Hollik:2012rc}, where it was investigated the NLO corrections to squark-squark pair production due to strong processes and to the exchange of a gluino in the t-channel. This reference finds $K \approx 1.4$, hence in our toy model, where the exchanged fermion in the t-channel is a colour singlet and not a colour octet, we expect a smaller impact of the NLO QCD corrections and accordingly a $K$ factor not larger than this value.

In order to incorporate the effects of the NLO corrections to the total production cross section we parametrize the full cross section as 
\begin{equation}\label{eq:prod}
   \sigma = \sigma_{\rm QCD}^{\rm NLO+NLL} + K \times ( \sigma^{\rm LO}(f) -
\sigma^{\rm LO}(0) )\;,
\end{equation}
where $\sigma^{\rm LO}(f)$ denotes the leading-order cross section for a given coupling $f$, namely $\sigma^{\rm LO}(0)$ is the LO QCD contribution. We compute the LO using CalcHEP3.2 \cite{Pukhov:2004ca} with the CTEQ6 PDF set ~\cite{Pumplin:2002vw}. To estimate the uncertainty, we  vary $\sigma_{\rm QCD}^{\rm NLO+NLL}$ within the theoretical error given in~\cite{Kramer:2012bx,LHCxsec}, and the $K$-factor within the range $0.8-1.3$. Furthermore, we take $K=1$ as fiducial value. Given the known $K$-factors for similar processes, we believe that this represents a conservative choice.
Since QCD-mediated processes do not necessarily dominate the production of $\eta$, we performed a full Monte Carlo simulation of production and detector response in order to re-interpret experimental searches for simplified SUSY models within the scenario considered in this work.

\section{Re-interpretation of LHC constraints}\label{sec:LHC}

In order to derive collider constraints, we use the ATLAS search \cite{TheATLAScollaboration:2013fha}
for jets and missing transverse energy, based on ${\cal L}=20.3$\,fb$^{-1}$ of data collected at a center of mass
energy of $8$\,TeV. The search requires missing transverse energy $E_T^{\rm miss}>160$\,GeV, and transverse momentum
$p_T>130$\,GeV for the leading jet, as well as $p_T>60$\,GeV for subleading jets. The search is divided into 
various signal regions characterized by the number of hard jets (ranging from two to six), as well as a number of further
cuts, that are designed to suppress backgrounds (mainly diboson, Z/W+jets and $t\bar t$) as specified in \cite{TheATLAScollaboration:2013fha}.
The systematic uncertainties in the background rates are calibrated against four control regions.

The search does not find a significant excess above backgrounds and therefore presents $95\%$C.L. upper limits $S_{95}^{\rm obs}$ on the
number of signal events in each signal region. For a given model, these are related to the upper limit on the production cross section $\sigma$ via
\begin{equation}\label{eq:LHClimit}
  S_{95}^{\rm obs} = \sigma_{\rm vis} \times {\cal L} = \sigma \times \epsilon \times {\cal L}\;,
\end{equation}
where $\sigma_{\rm vis}=\sigma\times\epsilon$ is the visible cross section, and the efficiency $\epsilon=N_{\rm after\ cuts}/N_{\rm generated}$ gives the fraction of the number of events passing all cuts required by
a given signal region.

While the upper limits on the number of signal events in each
signal region are model-independent, their interpretation in terms of simplified supersymmetric models relies on the
corresponding efficiencies, which are model-dependent. For a large portion of the parameter space, the most relevant production
channel is $uu\to \eta\eta$, different from the simplified
supersymmetric model considered in \cite{TheATLAScollaboration:2013fha}. Furthermore, for moderate mass splittings
$m_\eta-m_{\chi} \lesssim {\cal O}(10^2\,$GeV$)$ additional hard jets emitted either from the initial state, the final state or
an intermediate particle in the diagrams shown in Fig.\,\ref{fig:LHC} contribute significantly to the visible cross section.
Potentially, this introduces a further source of model-dependence.

Therefore, we determined the appropriate efficiencies for the model discussed in the previous section by generating a large
number of events using MadGraph5 \cite{Alwall:2011uj} interfaced with the detector simulation Delphes (version 3.0.10) \cite{Ovyn:2009tx}. Furthermore, we generate hard events with up to two additional partons in the final state. The potential double-counting with initial- and final-state radiation generated in the hadronization process (for which we use Pythia8 \cite{Sjostrand:2007gs}) is   taken into account  by employing the MLM matching scheme,  taking the minimum kt jet measure between partons to be \texttt{xqcut}=$m_{\eta}/4$ and the jet measure cutoff used by Pythia \texttt{QCUT=xqcut}. We validated the analysis by considering the simplified supersymmetric model with squark and neutralino and reproduced the cut-flow reported in \cite{TheATLAScollaboration:2013fha} for the corresponding benchmark points (we typically find agreement at the sub-$10$\%
level, ranging up to $30$\% at most). 

\begin{figure}
\hspace*{-0.6cm}
\begin{tabular}{cc}
 \includegraphics[width=0.5\textwidth]{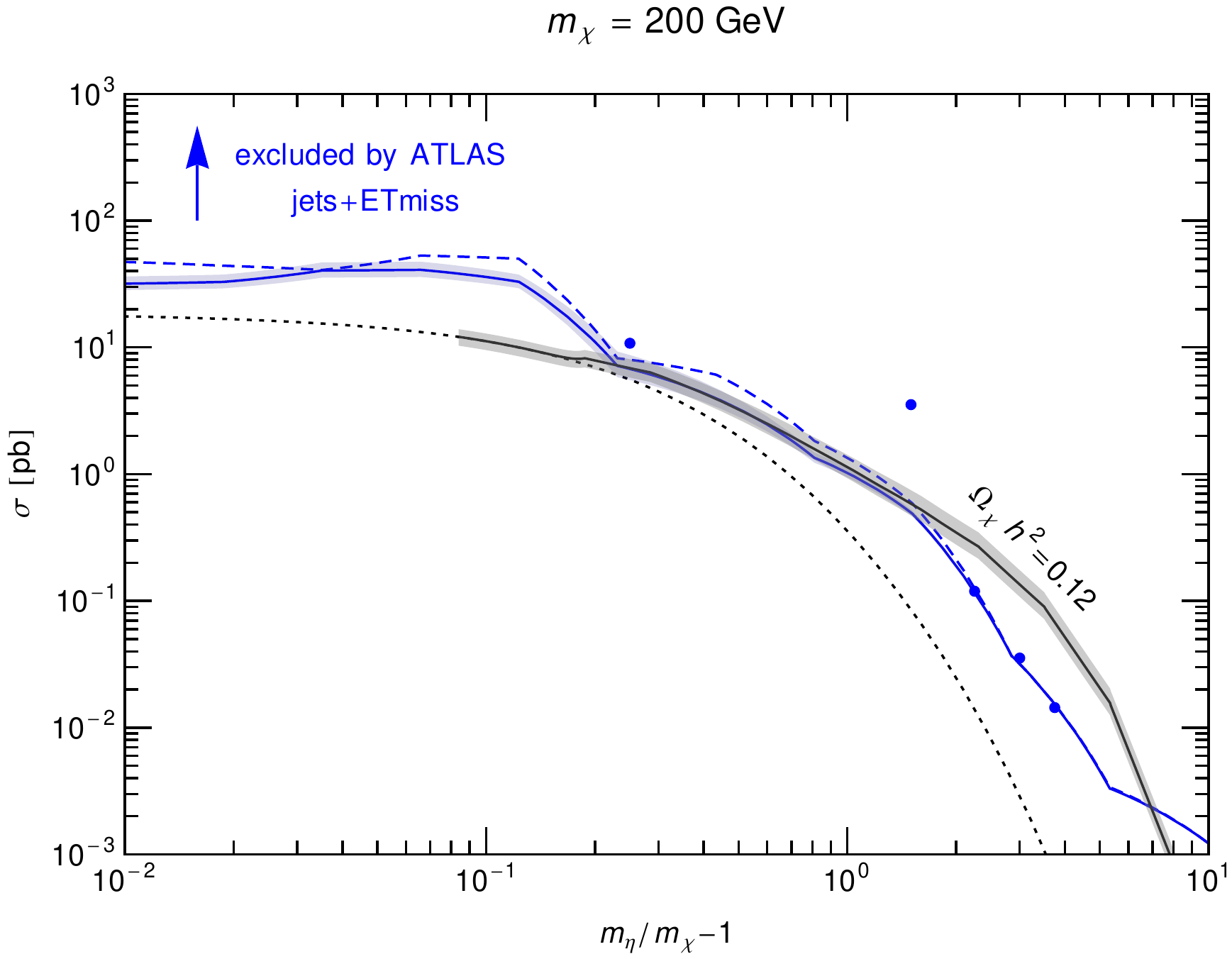} &
 \includegraphics[width=0.5\textwidth]{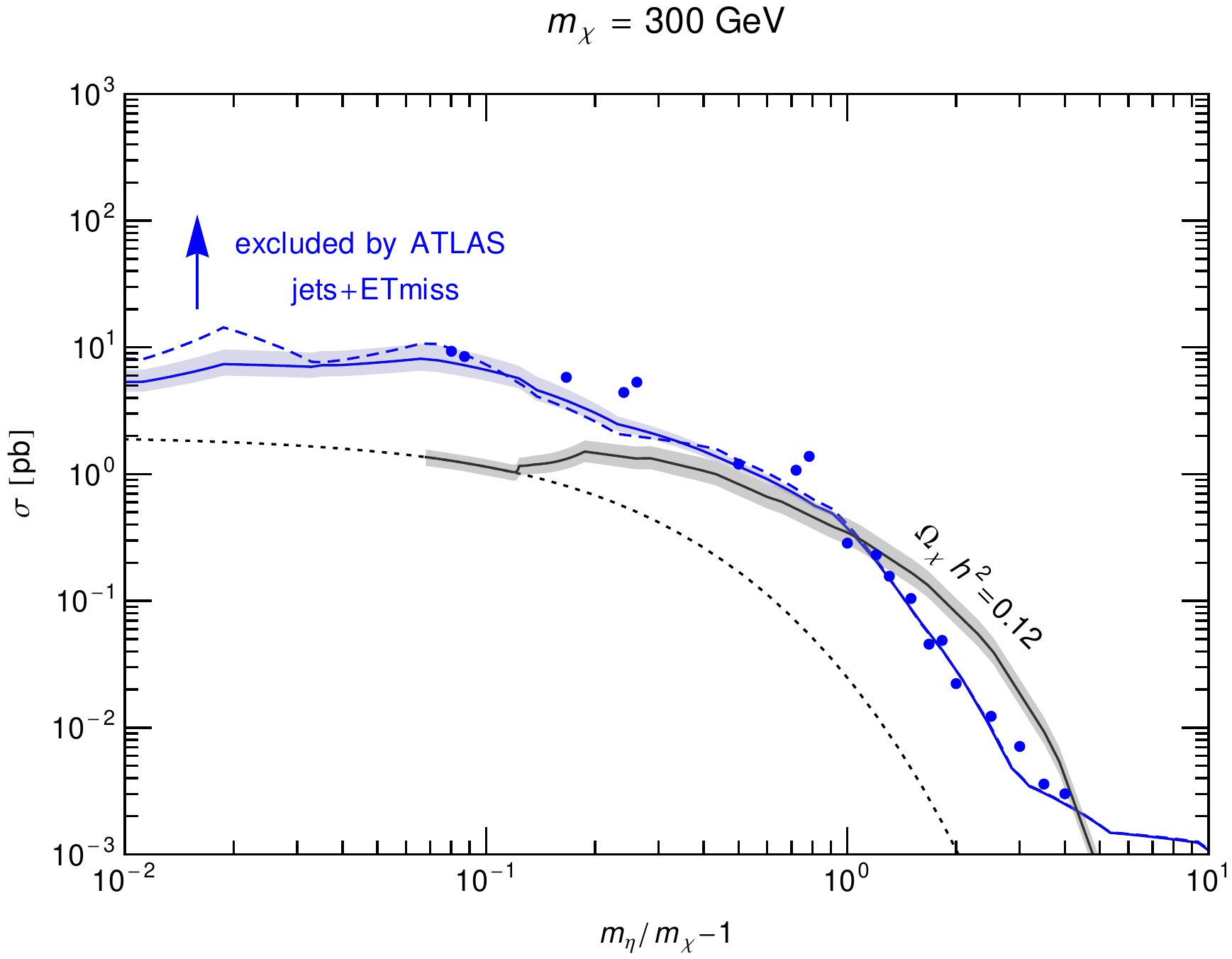}
 \\
 \includegraphics[width=0.5\textwidth]{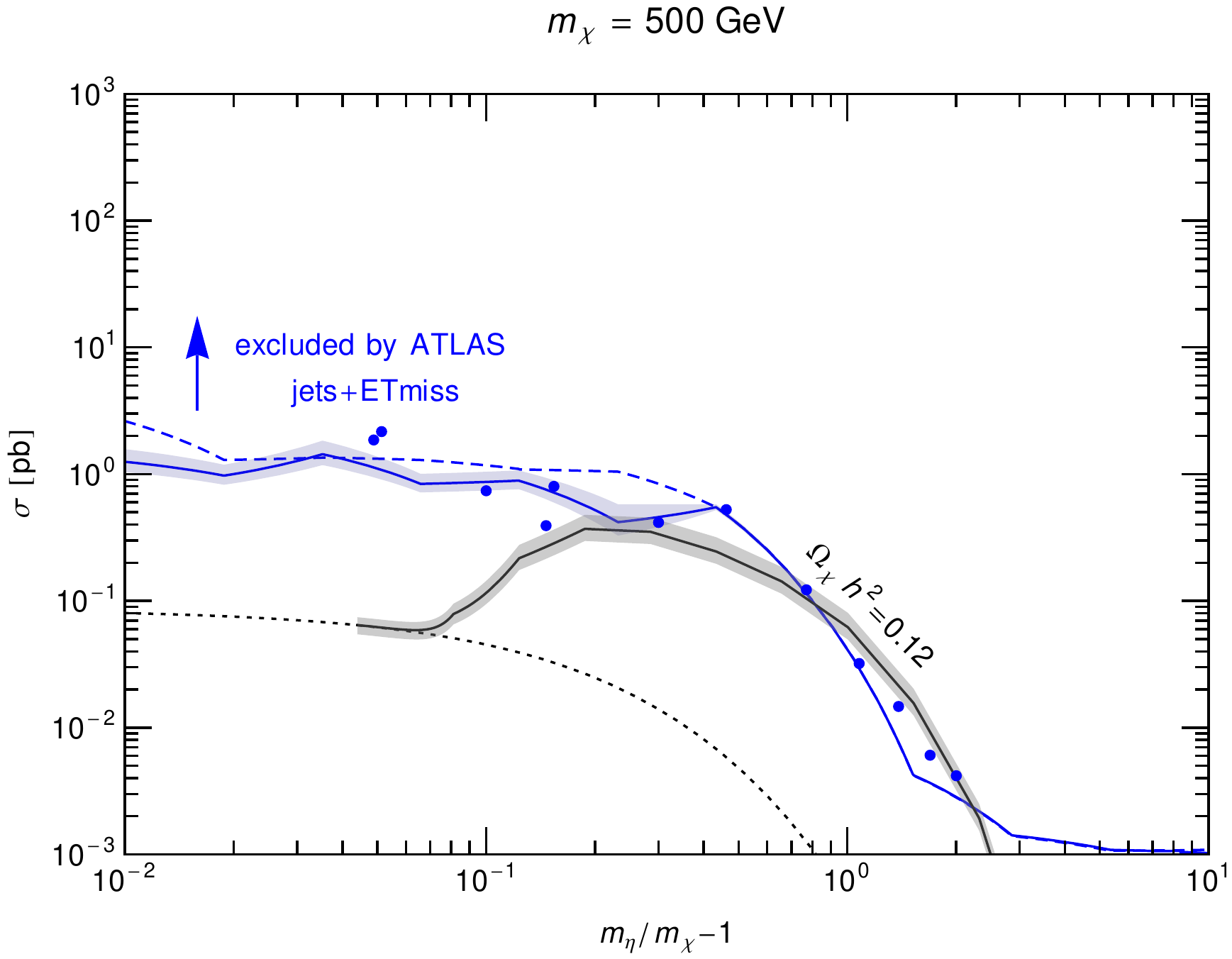} &
 \includegraphics[width=0.5\textwidth]{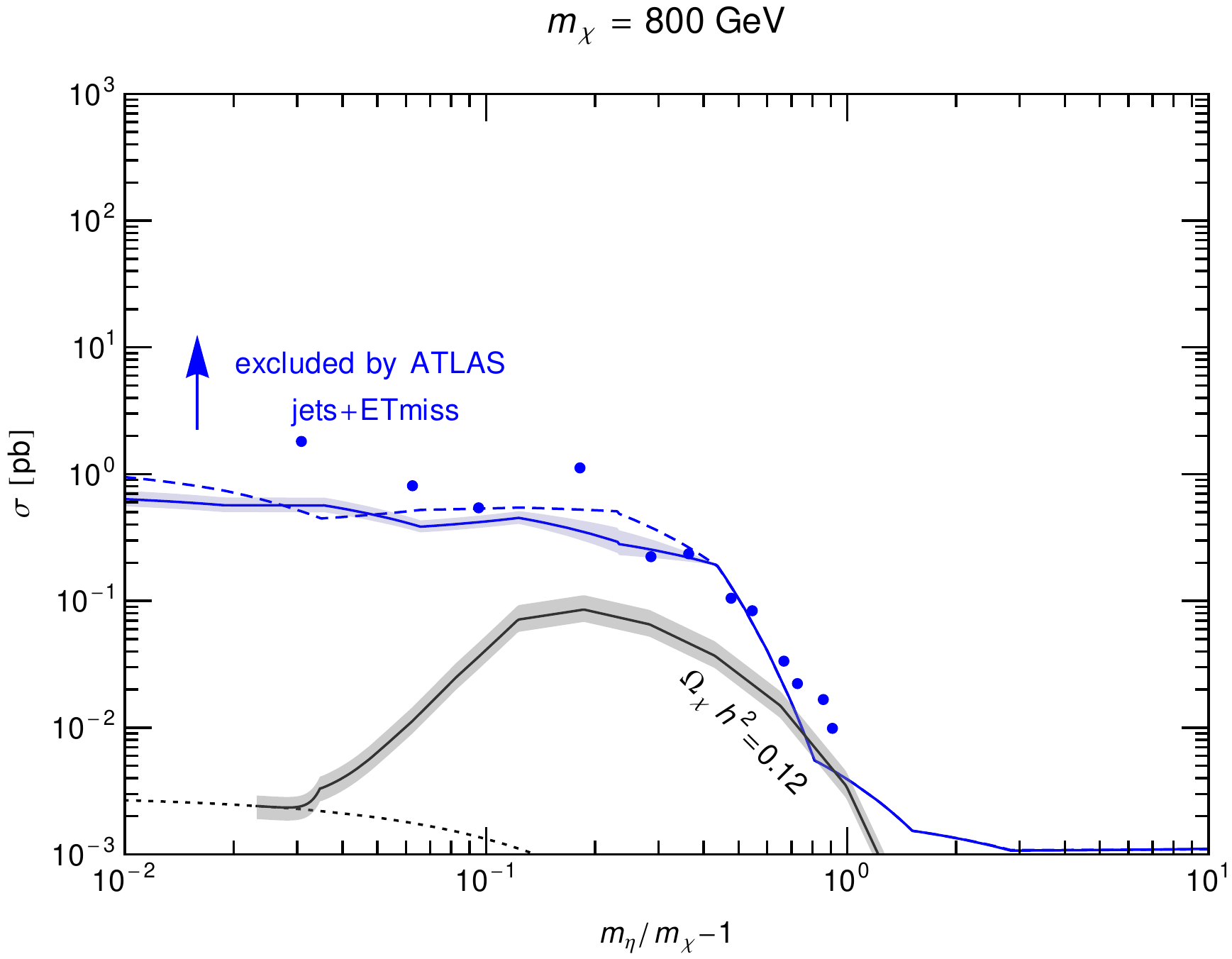}
\end{tabular}
 \caption{\label{fig:SplittingVsCrossSection} Upper limit on the production cross section from the ATLAS search \cite{TheATLAScollaboration:2013fha} for jets and missing
  transverse energy (blue) as a function of the mass splitting between the dark matter and mediator particle. The four panels correspond to $m_{\chi}=200, 300, 500, 800$\,GeV. The black line shows the expected cross section for a thermal WIMP. The black dotted line is the production cross section arising from QCD interactions only (i.e. for $f\to 0$). The blue dashed line corresponds to the limit one would obtain when including only one additional ISR/FSR jet in the matching. For comparison, the blue dots mark the upper limit given by ATLAS \cite{TheATLAScollaboration:2013fha} for a simplified supersymmetric model containing squarks and neutralino.}
\end{figure}

We compute the efficiencies for all signal regions containing up to four jets by generating a large number of events in a two-dimensional grid of parameter points for $m_{\chi}, m_\eta/m_{\chi}$, while setting the coupling $f=f_{\rm th}(m_\eta,m_{\chi})$ to the value expected for a thermal relic. For each point in parameter space, we then derive an upper limit on the production cross section via Eq.\,(\ref{eq:LHClimit}). We select the signal region with the best expected sensitivity $\epsilon\times S_{95}^{\rm exp}$, where the latter are taken from \cite{TheATLAScollaboration:2013fha}. The computation of efficiencies can be affected by uncertainties related to the matching procedure. To estimate this uncertainty we varied the matching scales \texttt{xqcut} and \texttt{QCUT} within a range of a factor of two, and find changes in the efficiencies below $\sim 30\%$. In addition, the efficiencies can be affected considerably by statistical uncertainties related to the finite number of 
generated events. This is critical in particular for very small masses $m_{\chi}$ and small splittings $m_\eta/m_{\chi}$, and for the signal regions containing three or four jets. In order to reduce this uncertainty as far as possible, we used up to $N_{\rm generated}\simeq 1.2\cdot 10^6$ events. Nevertheless, for a given set of masses, we exclude all signal regions for which the statistical $1\sigma$ error of the efficiency is above $30$\%. The corresponding uncertainty of the upper limit on the production cross section is shown as blue band in Fig.\,\ref{fig:SplittingVsCrossSection}. 

\begin{figure}
\begin{tabular}{cc}
 \includegraphics[width=0.47\textwidth]{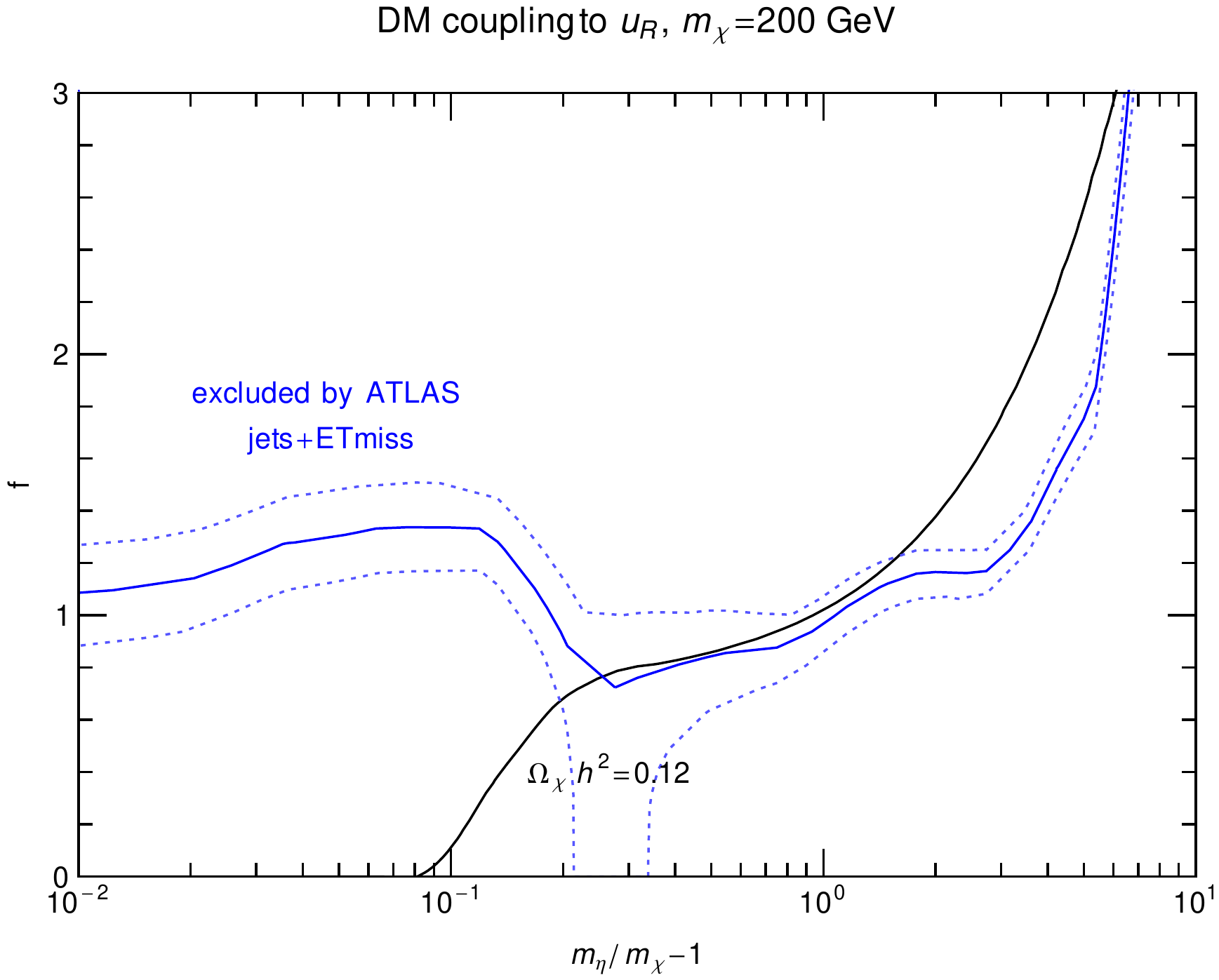} &
 \includegraphics[width=0.47\textwidth]{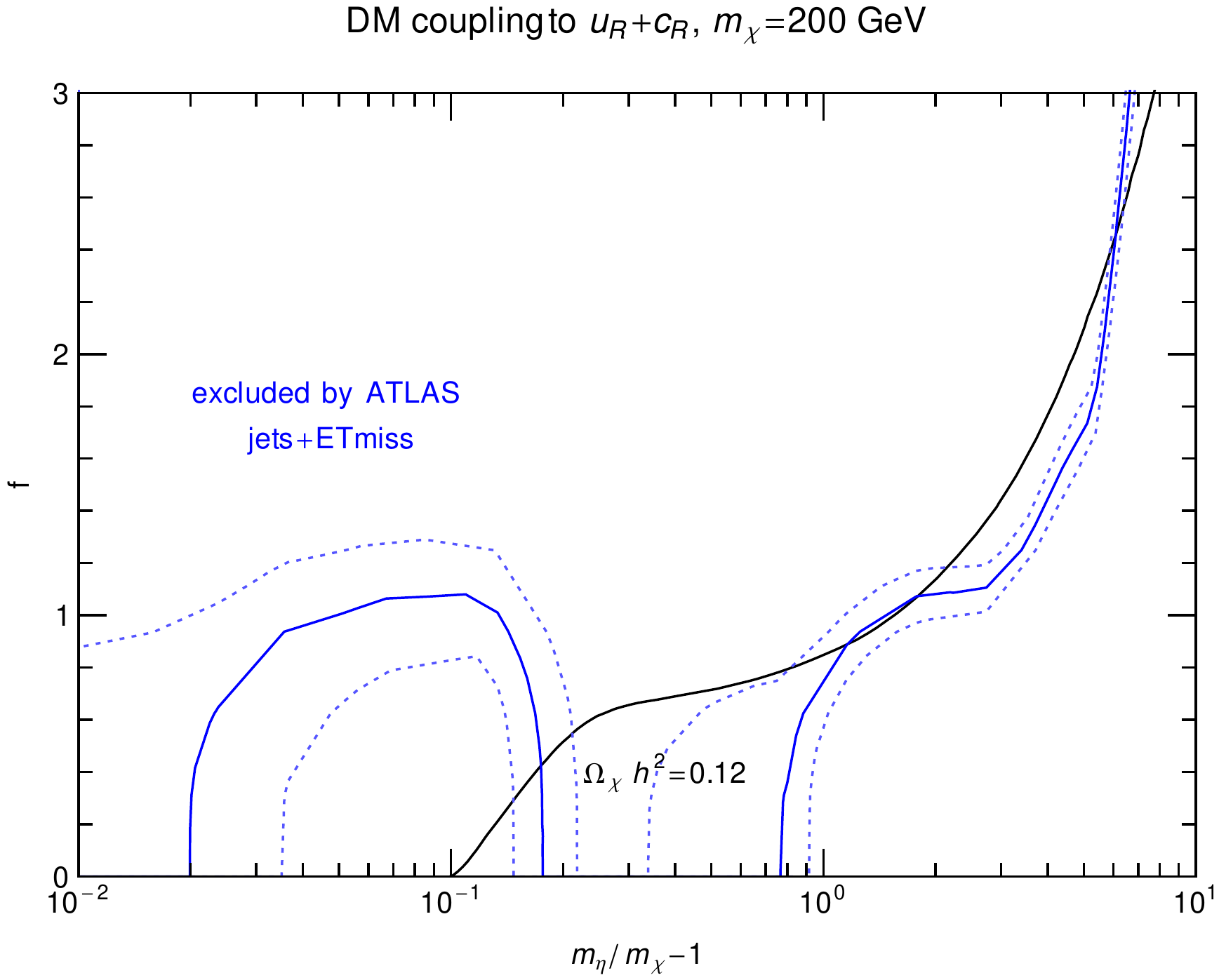}
 \\
 \includegraphics[width=0.47\textwidth]{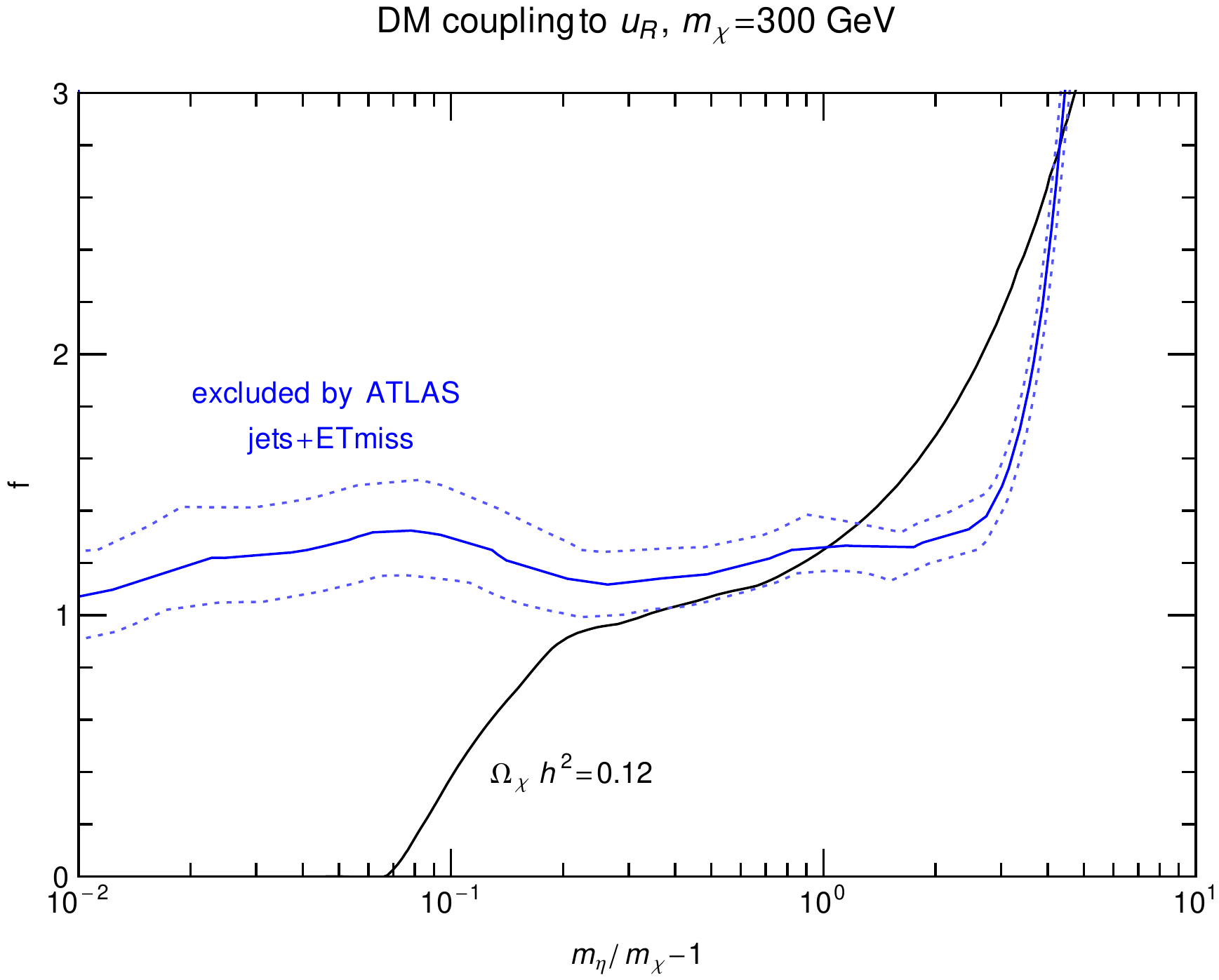} &
 \includegraphics[width=0.47\textwidth]{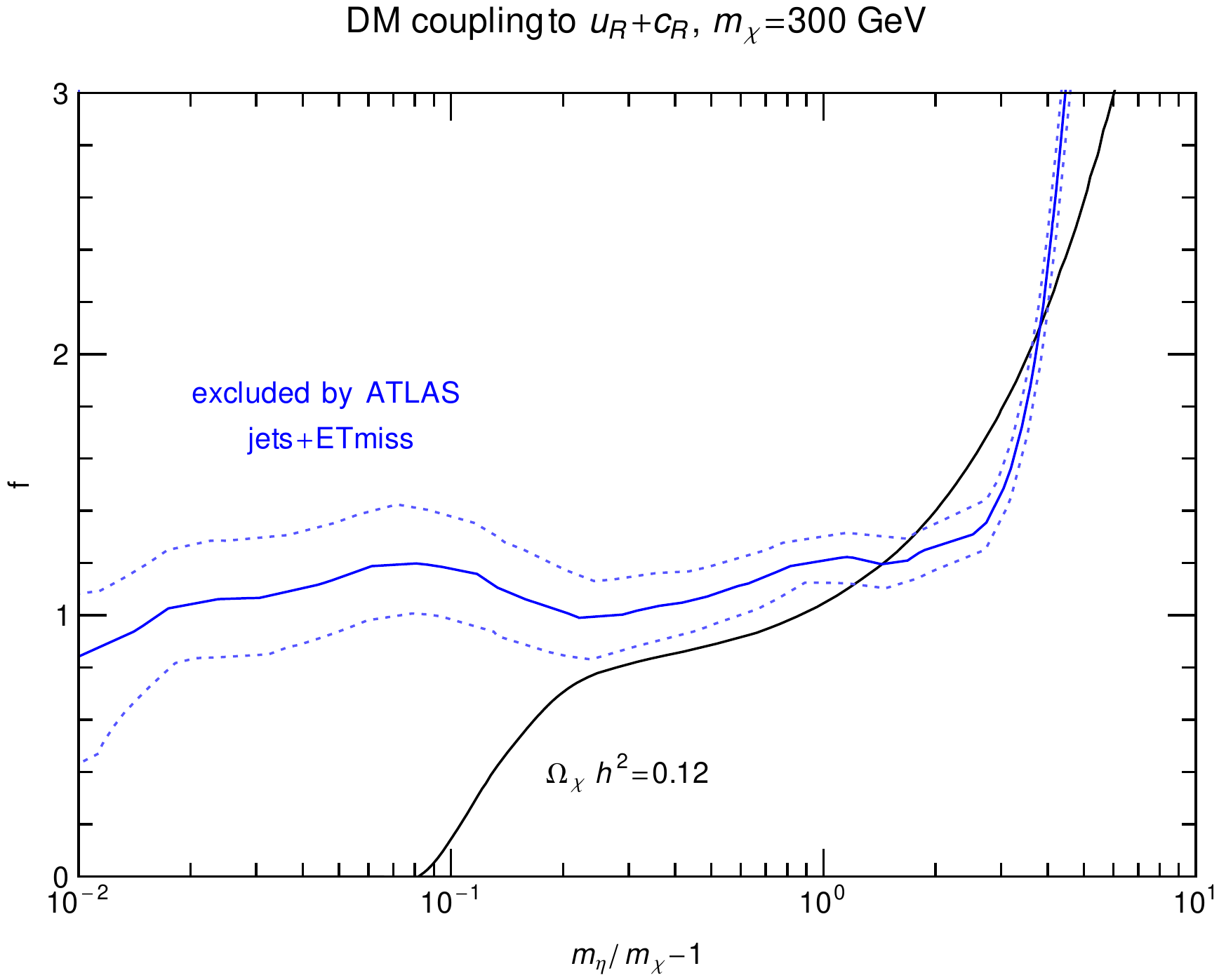}
 \\
 \includegraphics[width=0.47\textwidth]{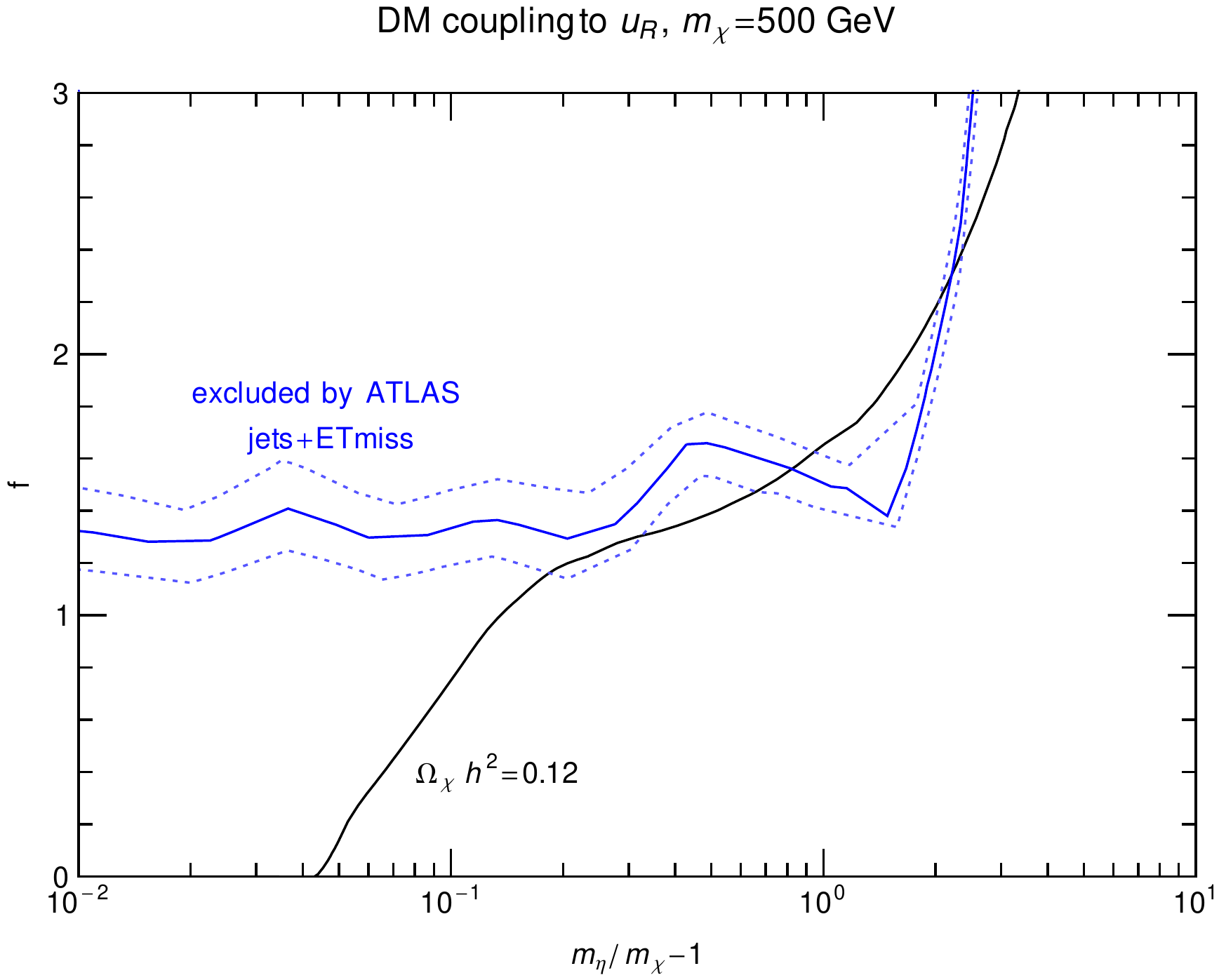} &
 \includegraphics[width=0.47\textwidth]{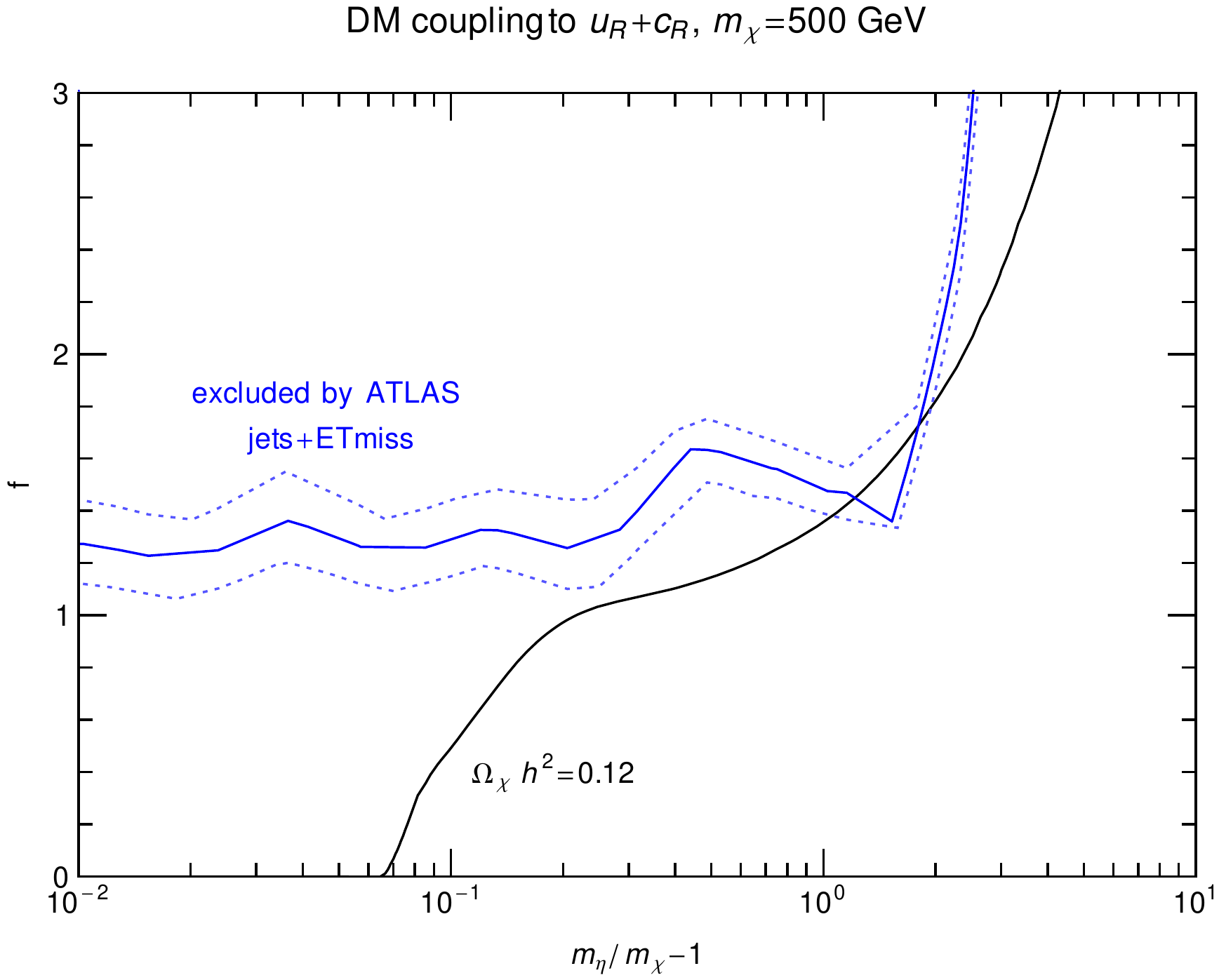}
\end{tabular}
 \caption{\label{fig:SplittingVsf} Upper limit on the coupling $f$ inferred from the ATLAS search \cite{TheATLAScollaboration:2013fha} for jets and missing
  transverse energy (blue) as a function of the mass splitting between the dark matter and mediator particle, for $m_{\chi}=200, 300, 500$\,GeV. The left panel corresponds to the case of a single mediator coupling to $u_R$, and the right to two mass-degenerate mediators coupling to $u_R$ and $c_R$, respectively. The blue dotted lines indicate the uncertainty (see text for details), and the black line corresponds to a thermally produced WIMP.}
\end{figure}

\section{Results}\label{sec:results}

In this section we present and discuss our main results obtained from re-interpreting the ATLAS search \cite{TheATLAScollaboration:2013fha} for jets and missing transverse energy for the dark matter model described in section~\ref{sec:model}, and compare the resulting constraints to those from direct and indirect searches. We mostly focus on the case of a single coloured mediator that couples to the $u_R$ quark for definiteness, but also consider the case of two mass-degenerate mediators that couple to $u_R$ and $c_R$, respectively.

As expected, the upper limit that can be placed on the production cross section depends strongly on the mass splitting $\delta\equiv m_\eta/m_{\chi}-1$ between the mediator $\eta$ and the dark matter particle $\chi$, see Fig.\,\ref{fig:SplittingVsCrossSection}. For $\delta \gg 1$, cross sections down to $\sim 1$\,fb can be excluded, the most sensitive signal regions being those with three and four jet  final states (more precisely Bt and Ct in the notation of \cite{TheATLAScollaboration:2013fha}) due to a better background suppression. At intermediate splittings $\delta\sim{\cal O}(1)$, the limit weakens by several orders of magnitude due to a loss in efficiency. Here the signal regions  with looser requirements, i.e. more statistics, tend to become more important (Bm and Cm). For very small mass splittings $\delta \ll 1$, the limit on the cross section reaches a plateau at $0.5-50$pb, depending on the mass. For small splittings the signal regions with two jets become important (Al and Am). This is due to a loss of efficiency in the $\geq 3$-jet signal regions, since the jets produced in the decay $\eta\to\chi q$ become too soft. Instead, the dominant contribution comes from additional hard jets radiated from initial, final or intermediate states. Consequently, when one would include only one additional jet in the matrix elements (our default is two), the exclusion limits would be weakened in this regime (blue dashed lines in Fig.\,\ref{fig:SplittingVsCrossSection}).

For comparison, we also show the production cross section that is expected for a thermal relic in Fig.\,\ref{fig:SplittingVsCrossSection} (black lines). The black dotted line is the contribution mediated by the strong interaction (first five diagrams in Fig.\,\ref{fig:LHC}). The additional production channel via t-channel exchange of $\chi$ (last two diagrams in Fig.\,\ref{fig:LHC}) typically yields the dominant contribution (especially $uu\to\eta\eta$) for mass splitting $\delta\gtrsim 0.1$. In general, their size depends on the coupling $f$.
For obtaining the black line in Fig.\,\ref{fig:SplittingVsCrossSection}, $f$ has been fixed by the requirement of producing the observed dark matter abundance via thermal freeze-out. For small mass splittings $\delta\lesssim 0.1$, coannihilation channels are important, such that rather small values of $f$ are sufficient to obtain the observed abundance. In turn, this means that the expected production cross section is suppressed in this regime. The grey band in Fig.\,\ref{fig:SplittingVsCrossSection} indicates the estimated uncertainty of the production cross section as discussed in Sec.\,\ref{sec:prodLHC}.

Instead of fixing the coupling $f$, one may also treat it as a free parameter and determine an upper limit by requiring that the production cross section remains below the experimental upper limit. As can be seen in Fig.\,\ref{fig:SplittingVsf}, the largest allowed values of $f$
are of ${\cal O}(1)$ for $\delta \lesssim 2-3$, and quickly grow for larger splittings. The blue dotted lines in Fig.\,\ref{fig:SplittingVsf} are obtained by varying both the exclusion cross section as well as the production cross section within the uncertainty ranges discussed in Sec.\,\ref{sec:LHC} and Sec.\,\ref{sec:prodLHC}, respectively. Note that, whenever the QCD contribution to the production lies above the exclusion limit on its own, the upper limit
on the coupling $f$ formally approaches zero. This means that this set of masses is excluded independently of the value of $f$.

The left panel of Fig.\,\ref{fig:SplittingVsf} shows the
case of a single mediator that couples to $u_R$, and the right panel corresponds to two mediators coupling to $u_R$ and $c_R$. The constraints are comparable, and the largest differences arise for small dark matter masses $m_{\chi}\lesssim 200$\,GeV. The reason is that, in this region, the contribution to the production cross section arising from gluon initial states is important. This contribution doubles when considering two mediators, because it is flavour-insensitive. On the other hand, the production cross section is affected very little for the channels that depend on $f$, because these are typically dominated by $uu\to \eta\eta$. This explains why the corresponding limits are very similar to the case of a single mediator. However, the value of the coupling $f$ required for producing the observed dark matter abundance via freeze-out is sensitive to the number $N$ of mediators. Outside of the coannihilation region, $\Omega_\chi \propto N\times f^4$, i.e. $f_{\rm th}\propto N^{-1/4}$ (for our numerical results we computed the relic density with micrOMEGAs in all cases, to take coannihilation effects into account).

\subsection{Comparison with direct detection}

\begin{figure}
\begin{tabular}{cc}
 \includegraphics[width=0.48\textwidth]{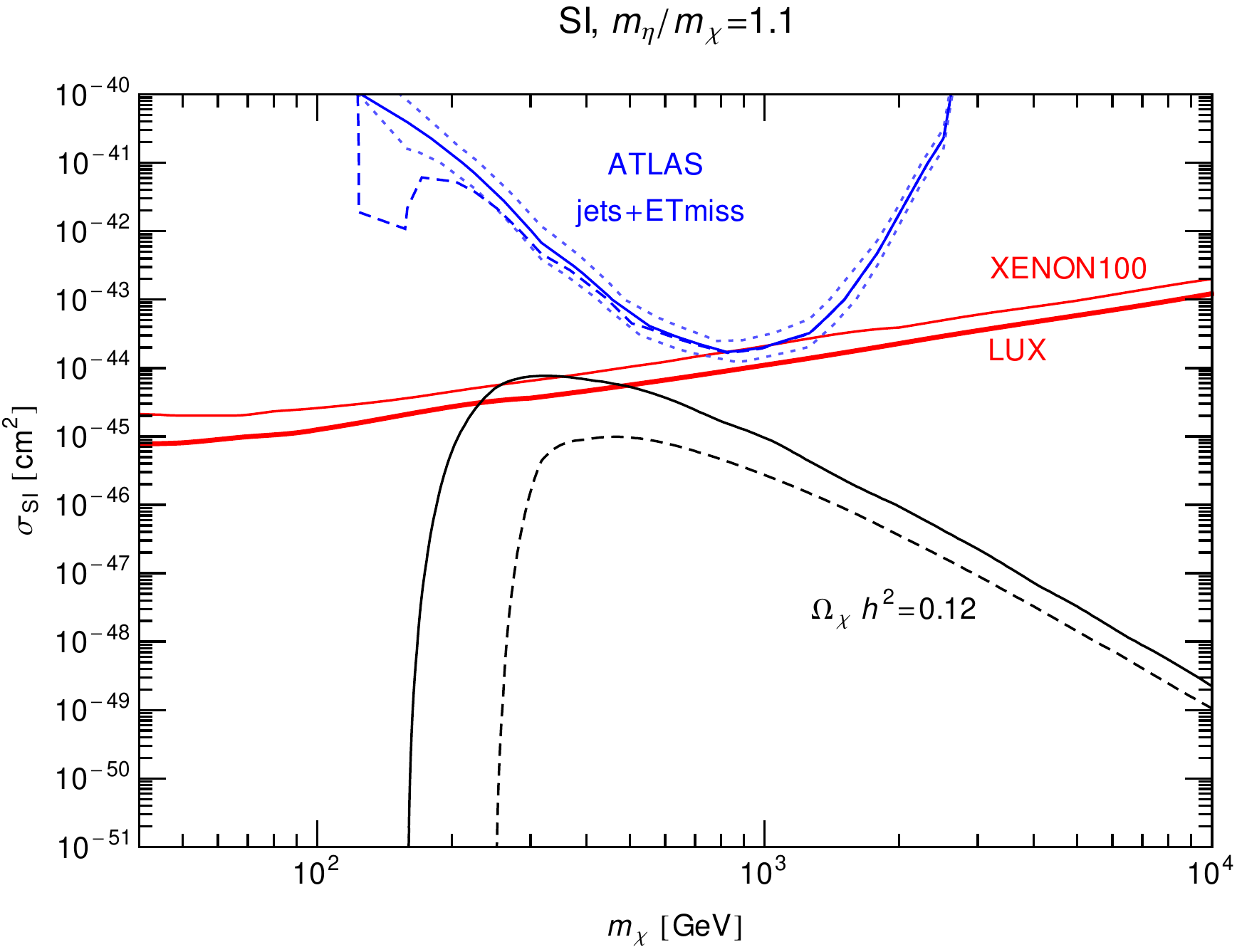} &
 \includegraphics[width=0.48\textwidth]{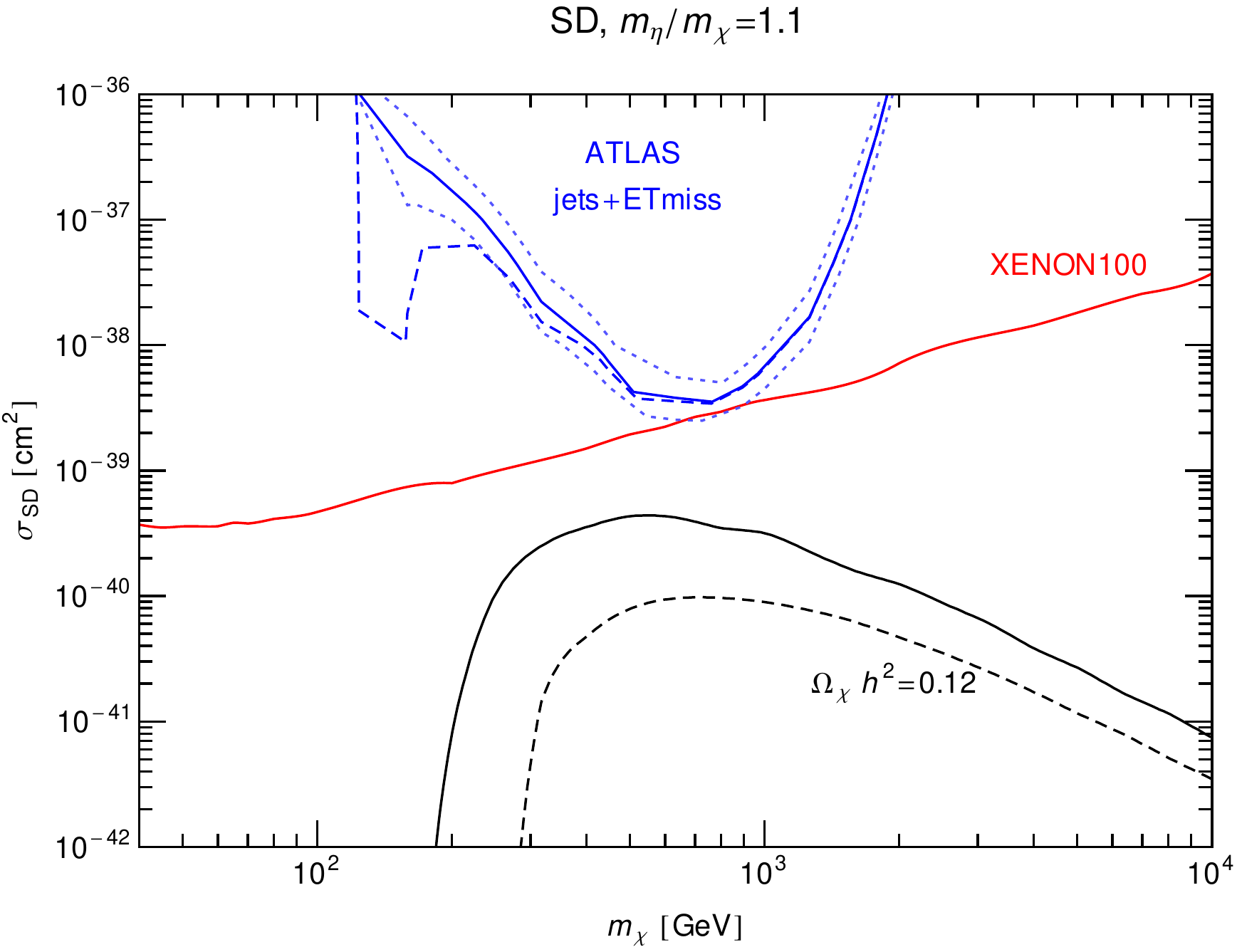}
 \\
 \includegraphics[width=0.48\textwidth]{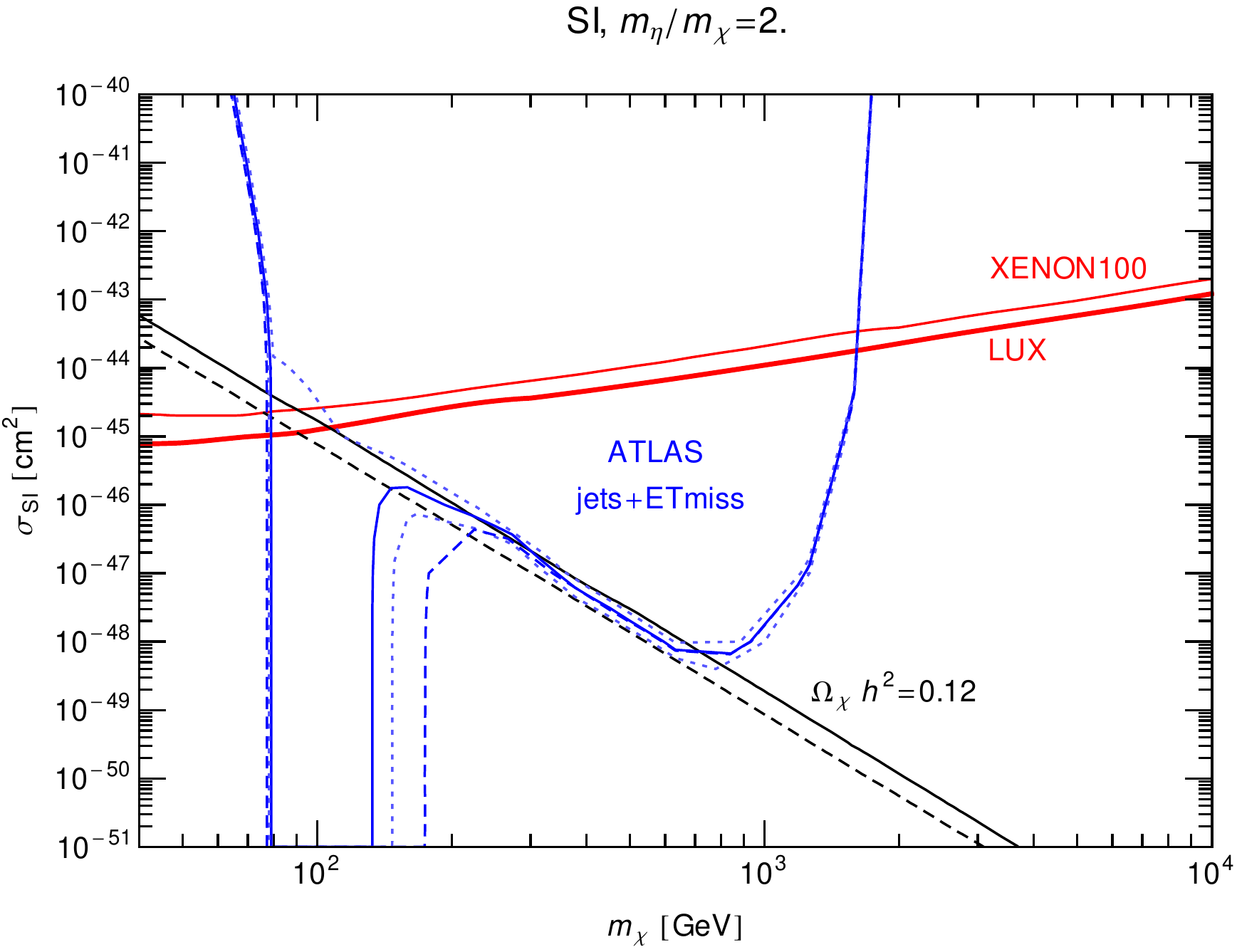} &
 \includegraphics[width=0.48\textwidth]{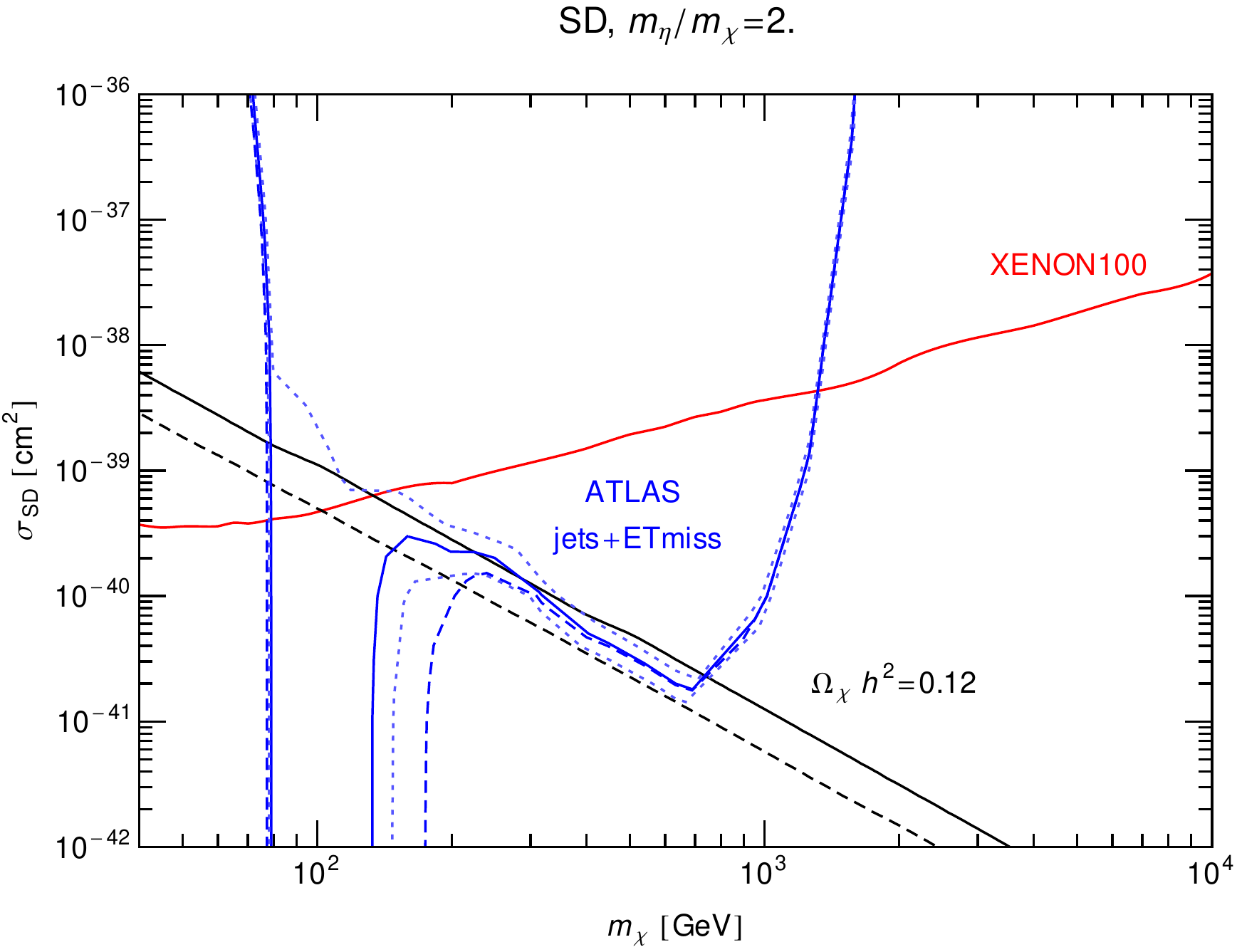}
 \\
 \includegraphics[width=0.48\textwidth]{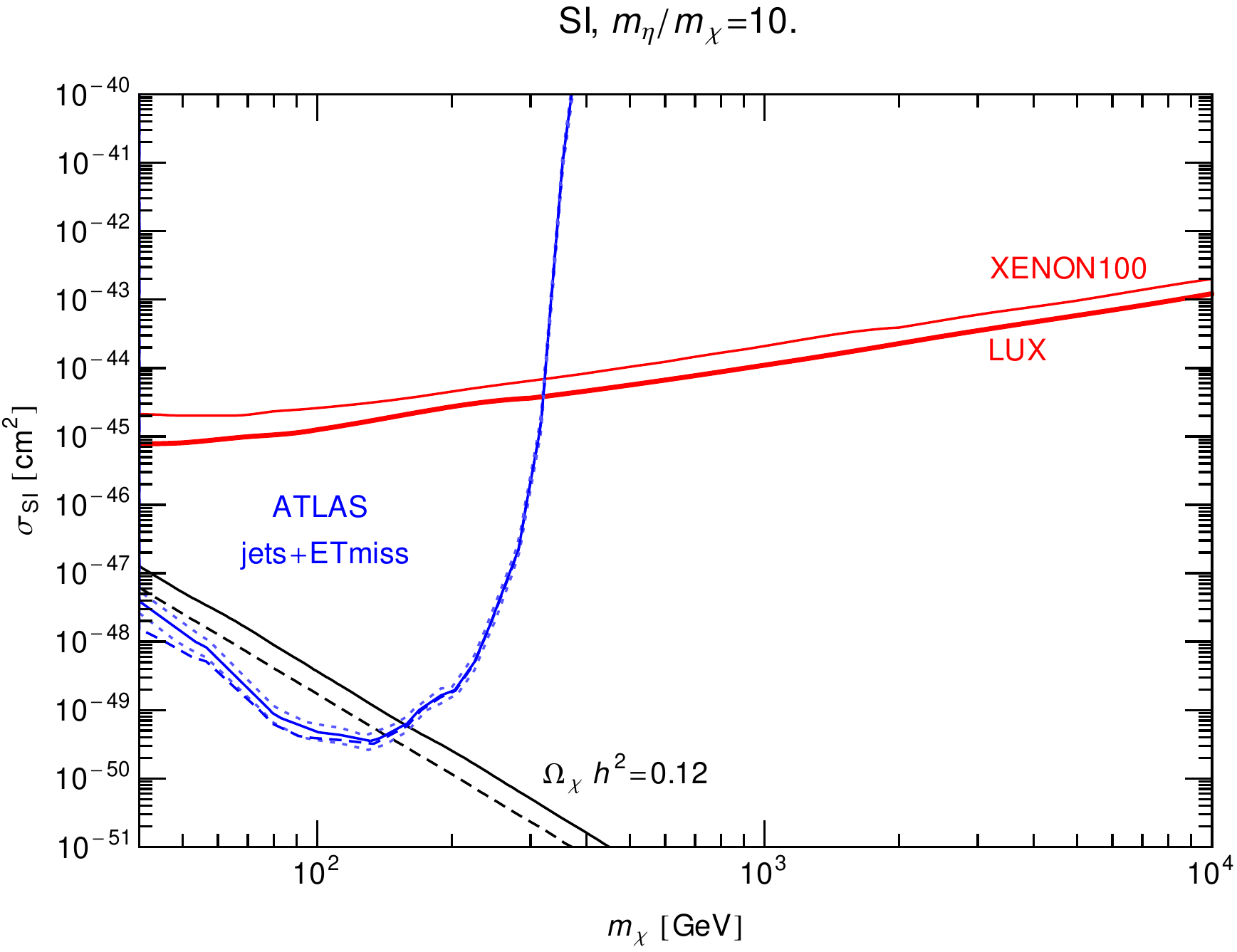} &
 \includegraphics[width=0.48\textwidth]{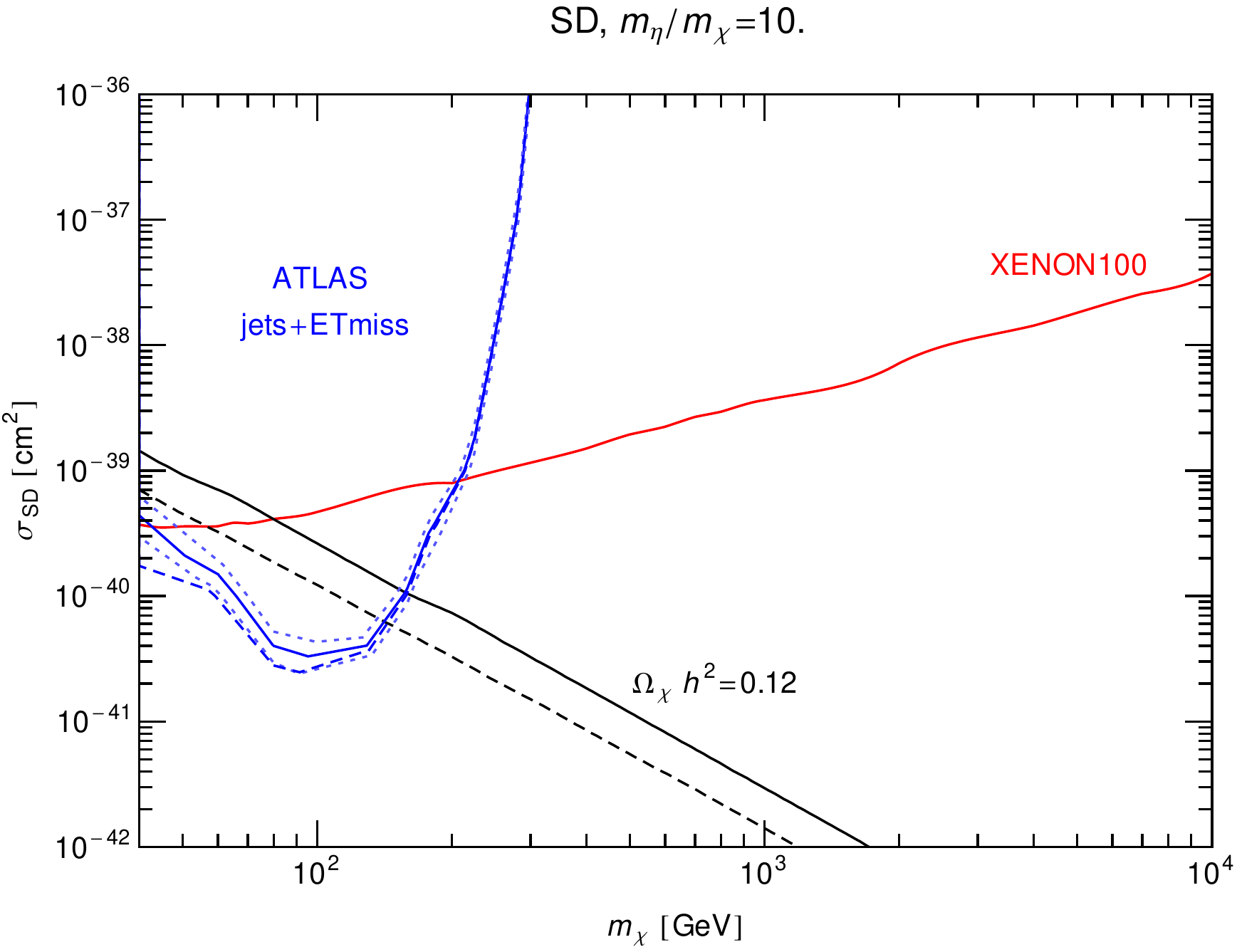}
\end{tabular}
 \caption{\label{fig:DD} Comparison of constraints inferred from the ATLAS search~\cite{TheATLAScollaboration:2013fha} for jets and missing energy with the spin independent (-dependent) scattering cross section off protons (neutrons), shown in the left (right) panel. The rows correspond to mass ratios $m_\eta/m_{\chi}=1.1,2,10$ between the dark matter and mediator mass. Solid lines correspond to the case where dark matter couples to $u_R$, and dashed where it also couples to $c_R$. The blue dotted lines indicate the uncertainty of the collider constraint for the case of $u_R$-coupling (see text for details).}
\end{figure}

The upper limits on the coupling $f$ can be translated into limits on the spin independent and spin dependent scattering
cross section. The corresponding constraints are shown in Fig.\,\ref{fig:DD}, together with upper limits from 
XENON100 \cite{Aprile:2012nq,Aprile:2013doa} and LUX \cite{Akerib:2013tjd}. For small mass splitting $\delta\ll 1$, the direct detection cross sections are resonantly enhanced, while the collider limits are weakened as discussed above. On the other hand, for $\delta={\cal O}(1)$, the collider search is very effective, while the direct detection cross section is suppressed for Majorana dark matter with chiral couplings, as discussed in Sec.\,\ref{sec:DD}. Consequently, when converted into the direct detection cross section, the ATLAS limits can be stronger by one to several orders of magnitude than current bounds from direct detection experiments for masses in the range $m_{\chi}=10^2-10^3$\,GeV.
For masses around $100$\,GeV and $\delta={\cal O}(1)$, the ATLAS constraint is strong enough to exclude even the QCD contribution to the production cross section at LHC. This translates into the dip in the constraint visible in the middle row of Fig.\,\ref{fig:DD}. Note, however, that while in general collider uncertainties only have a moderate impact,
 the upper limit is considerably affected in this range (see blue dotted lines
in Fig.\,\ref{fig:DD}). For comparison, the ATLAS constraint for $u_R/c_R$ mediators is also shown by the blue dashed lines in Fig.\,\ref{fig:DD}. The cross section expected for a thermal relic is also shown as black solid (dashed) line for $u_R$ ($u_R/c_R$) mediators.

\subsection{Comparison with indirect detection}

\begin{figure}
\begin{center}
 \includegraphics[width=0.7\textwidth]{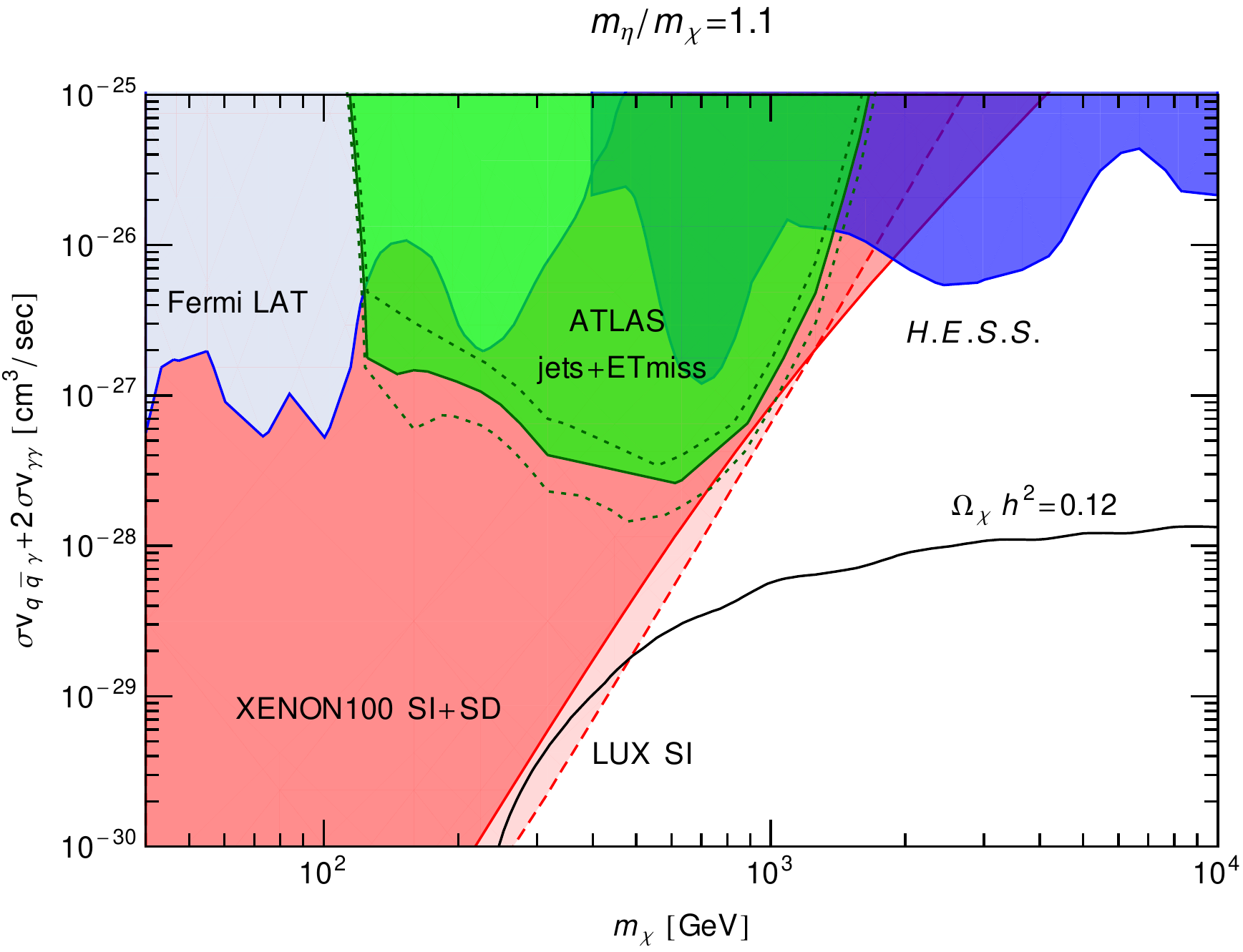}
 \\[3ex]
 \includegraphics[width=0.7\textwidth]{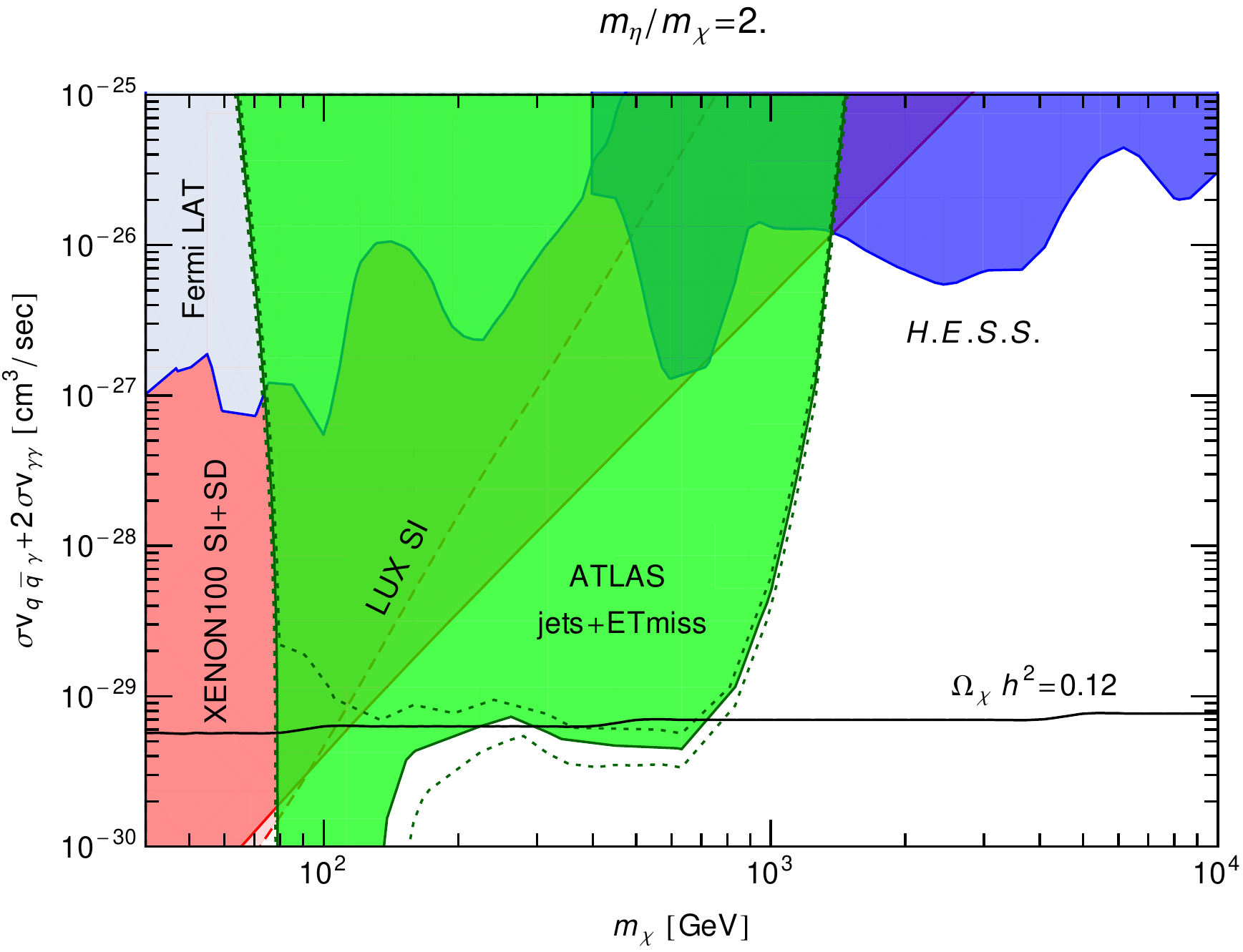} 
\end{center}
 \caption{\label{fig:mDMvsSigv} Comparison of constraints on the annihilation cross section obtained from searches for spectral features by the Fermi-LAT \cite{Bringmann:2012vr} and H.E.S.S. \cite{Abramowski:2013ax} ({\it cf}. \cite{Garny:2013ama}), with constraints inferred from collider searches for jets and missing energy by ATLAS  \cite{TheATLAScollaboration:2013fha}, as well as direct detection limits from XENON100 \cite{Aprile:2012nq} and LUX \cite{Akerib:2013tjd}. The black line corresponds to a thermal WIMP, and the dotted lines indicate the uncertainty of the collider constraint, as discussed before. Note that the results for $u_R/c_R$ mediator are very similar, and are therefore not shown. }
\end{figure}

\begin{figure}
\begin{center}
 \includegraphics[width=0.95\textwidth]{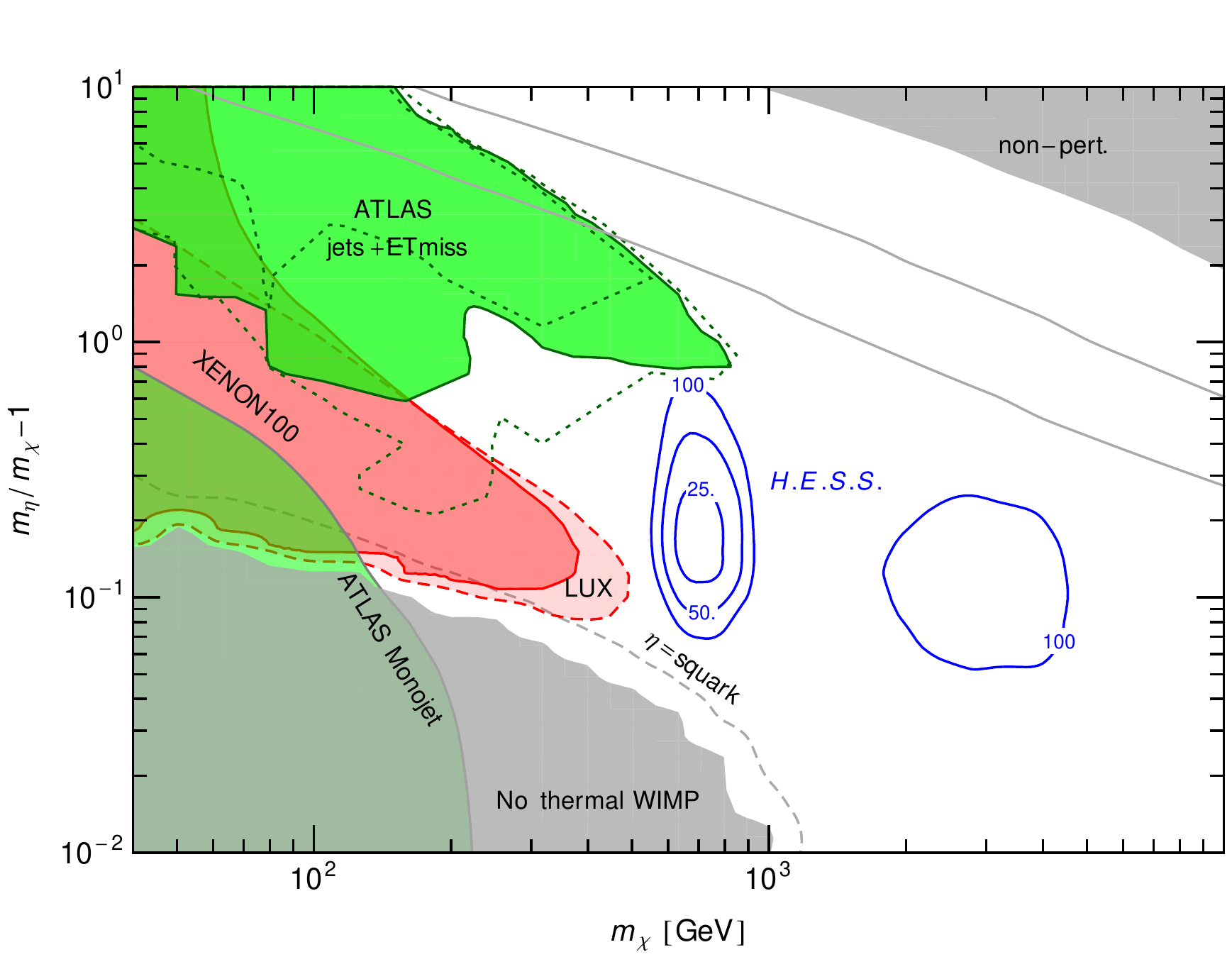}
\end{center}
 \caption{\label{fig:mDMvsSplitting_lhc_thermalCoupling} Constraints on thermally produced WIMP dark matter with a coloured mediator particle $\eta$. The green region is excluded at $95\%$C.L. by the ATLAS search \cite{TheATLAScollaboration:2013fha} for jets and missing transverse energy. For comparison, the red shaded area is excluded by direct searches. The blue lines indicate the regions excluded by the search for an internal bremsstrahlung feature in the gamma-ray spectrum from the central galactic halo measured by H.E.S.S. \cite{Abramowski:2013ax}, assuming a boost factor $25, 50$ or $100$, respectively \cite{Garny:2013ama}. Within the grey shaded region in the lower left corner, thermal production cannot make up for the whole dark matter abundance due to efficient coannihilations. Within the upper right corner, non-perturbatively large values of $f\gtrsim 10$ would be required. Below the upper(lower) gray line $\Gamma_{\eta}/m_\eta<0.5(0.1)$. The gray dashed line indicates the masses for which the coupling of the mediator equals the one of a squark.}
\end{figure}

One of the most interesting features of the dark matter model discussed here is the presence of a sharp spectral feature in the dark matter annihilation spectrum. It arises mainly from internal bremsstrahlung $\chi\chi\to q\bar q\gamma$ for $\delta \lesssim {\cal O}(1)$, while for $\delta\gtrsim {\cal O}(1)$, also the gamma-ray line resulting from the loop-induced process $\chi\chi\to \gamma\gamma$ gives a significant contribution. In Fig.\,\ref{fig:mDMvsSigv}, we compare constraints on $\sigma v_{q\bar q\gamma}+2\sigma v_{\gamma\gamma}$ from gamma-ray observations of the central galactic halo by Fermi-LAT \cite{Bringmann:2012vr} and H.E.S.S.  \cite{Abramowski:2013ax} (blue shaded regions) with those inferred from direct detection \cite{Garny:2013ama} and from the ATLAS search \cite{TheATLAScollaboration:2013fha} considered here. As expected, for small mass splitting, the region excluded by the LHC search (green region) is less constraining than XENON100 and LUX (red region). However, for mass splitting of order one, the ATLAS search severely constrains the possibility to observe a spectral feature arising from dark matter with a coloured mediator below ${\cal O}(1)$\,TeV energies. 

\subsection{Exclusion for a thermal relic}

By requiring that thermal freeze-out yields a relic abundance that coincides with the value measured by Planck~\cite{PlanckCollaboration2013}, it is possible to fix the coupling $f=f_{\rm th}(m_\eta,m_{\chi})$ between dark matter $\chi$, the mediator $\eta$ and the SM quarks for each set of masses. Under this assumption, the model has only two free parameters, which we take to be the dark matter mass $m_{\chi}$ and the mass splitting $\delta=m_\eta/m_{\chi}-1$. The collider limits considered here can be translated into an exclusion region, which we show in Fig.~\ref{fig:mDMvsSplitting_lhc_thermalCoupling} (green region). For mass splitting $\delta=1$, it reaches up to $m_{\chi}\sim 1$\,TeV. However, for smaller or larger mass splitting, much lighter masses remain allowed. On the one hand, for much larger splitting, the mediator $\eta$ becomes too heavy to be produced effectively. For much smaller splitting, on the other hand, two effects play a role: first, the collider search becomes less effective in this regime. Second, the coupling $f_{\rm th}$ gets very small due to efficient coannihilations. The combination of these effects also leads to the relatively large uncertainties in the exclusion region (green dotted lines), in particular as the thermal cross section and the LHC exclusion happen to exhibit a fairly similar dependence on the mass splitting  in certain regions of parameter space, see e.g. Fig.~\ref{fig:SplittingVsCrossSection}. For comparison, direct detection mostly probes a region with smaller mass splitting, due to the resonant enhancement of the nucleon scattering cross section for $\delta <1$ (red regions; note that for LUX only limits on spin independent scattering are available at present). The limits from indirect detection are currently not sensitive to the flux expected for a thermal relic, if the standard Einasto profile from \cite{Garny:2013ama} is adopted. However, if the flux is enhanced by a boost factor of order $25-100$, they probe the multi-TeV region (blue contour lines).
For comparison, we also show a constraint inferred from mono-jet searches for nearly degenerate particle spectra \cite{Dreiner:2012gx}, which is sensitive to very small splittings for low dark matter masses.

\section{Conclusions}\label{sec:conclusions}

The Large Hadron Collider offers a unique environment to search for dark matter particles with masses below $\sim 1 $ TeV through their possible production in partonic collisions. To optimize the search it is convenient to identify simplified models that characterize the signals of a larger class of dark matter models. In this paper we have focused on a model with Majorana dark matter particles that couple to the up-type quarks via one or several coloured mediators and which produces a signal consisting in two or more jets plus missing transverse energy, through the production and subsequent decay of the coloured scalar particles.

We have carefully analysed the production of coloured scalar particles at the LHC, considering not only the production via the strong gauge interaction, but also via the exchange of a dark matter particle in the t-channel. The latter production channel can be relevant and even dominant in some regions of the parameter space leading to the observed dark matter abundance via the thermal freeze-out of dark matter particles in the early Universe. More specifically, we have emphasized the importance of the partonic subprocess $u u \rightarrow \eta\eta$ mediated by a Majorana dark matter particle in the t-channel. Due to the enhancement of the rate by the square of the dark matter mass and due to the unsuppressed parton distribution function of up-quarks inside the proton, this process is the dominant production channel in large regions of the parameter space. Concretely, for large dark matter masses and a coloured scalar with comparable mass, we have found that the total production cross section of coloured scalars can be enhanced by more than two orders of magnitude compared to the production channels mediated by the strong interactions.

We have then derived limits on the parameters of the model employing the ATLAS search \cite{TheATLAScollaboration:2013fha} for jets and missing transverse energy, based on ${\cal L}=20.3$\,fb$^{-1}$ of data collected at a center of mass energy of $8$\,TeV. To re-interpret the analysis for the model considered here, we have computed the appropriate efficiencies for the relevant production channels, taking jet matching with two additional hard jets into account, for all signal regions containing two to four jets. Next, we have investigated the complementarity of the collider  limits with those from direct detection and indirect detection experiments. 

We have found that, for some regions of the parameter space of the model, the ATLAS searches imposes the strongest limits and rules out choices of parameters leading to the observed dark matter abundance via thermal production. For small mass splitting between the dark matter and the mediator, the collider limits are comparable to bounds from direct detection. However, if the mass splitting is of the same order as the dark matter mass, the ATLAS limits are considerably stronger than the latest bounds from XENON100 and LUX, reaching down to $\sigma_{SI}\sim 10^{-45}-10^{-48}$cm$^2$ for $200\,$GeV\,$\lesssim m_{\eta}\lesssim 2$\,TeV and $m_{DM}\lesssim m_{\eta}/2$. This is due to a relative suppression of the spin-independent scattering cross section for Majorana dark matter with chiral couplings, and the enhancement of the production at LHC described above. 

Searches for sharp spectral features at gamma-ray telescopes are fully complementary in the multi-TeV region. However, for $m_{DM}\lesssim 1$\,TeV direct detection and collider constraints in some cases even preclude the possibility of observing sharp spectral features at future gamma-ray telescopes for the standard choices of the astrophysical parameters. It is important to stress that these limits do not suffer from astrophysical uncertainties and are therefore very robust.  We have estimated uncertainties arising from the determination of efficiencies and from higher-order contributions to the production cross section, which are typically $\lesssim {\cal O}(50)\%$ but can be larger in particular cases. Lastly, we have also considered an extension of the model by extra coloured scalars, inspired by the particle content of the Minimal Supersymmetric Standard Model. We have found that our main conclusions still remain for the scenarios in agreement with the flavour physics experiments.

\section*{Acknowledgements}

We are grateful to Miguel Pato for earlier collaborations on which part of this work was based, and to Andreas Weiler for helpful discussions and for cross-checking efficiencies. This work has been partially supported by the DFG cluster of excellence ``Origin and Structure of the Universe'' and by the DFG Collaborative Research Center 676 ``Particles, Strings and the Early Universe''. S.V.~also acknowledges support from the DFG Graduiertenkolleg ``Particle Physics at the Energy Frontier of New Phenomena''.

\begin{appendix}

\section{Flavour constraints}
\label{App:flavour}

The interaction term of the dark matter particle and the coloured scalars with the right-handed quarks in general violates the $SU(3)_{u_R}$ flavour symmetry. Therefore,  it is necessary to check whether the stringent constraints arising from flavour physics are satisfied. In this Appendix we discuss how two well-known possibilities to suppress flavour-changing neutral currents, namely degeneracy or alignment, can be realized within the toy-model considered in this work.

\begin{figure}
\begin{center}
 \includegraphics[width=0.95\textwidth]{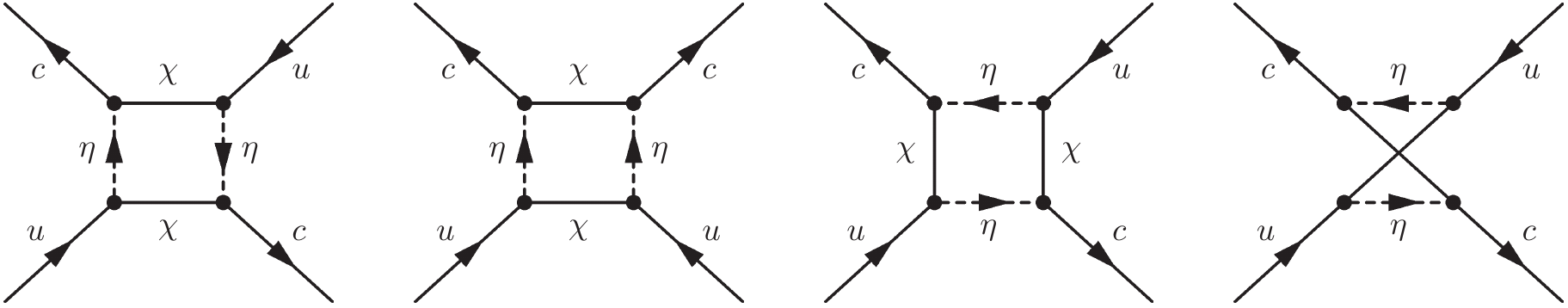}
\end{center}
 \caption{\label{fig:DDbar} Box diagrams giving rise to a non-standard contribution to $D-\bar D$ mixing.}
\end{figure}

Consider first the possibility of a single coloured scalar $\eta$, but allowing for arbitrary couplings $f_i$ to all right-handed quarks,
\begin{equation}
  {\cal L} = -f_i \bar u_{Ri} \chi \eta
\end{equation}
where $i=1,2,3$ corresponds to $u,c,t$. In this case the box diagram shown in fig.~\ref{fig:DDbar} gives a contribution to $D-\bar D$ mixing, which is  strongly constrained by the measured value of the $D$-meson mass splitting $\Delta m_D$ (note that there is no contribution to CP violation in presence of a single species $\eta$, such that constraints from $\epsilon_D$ do not apply). The box diagram gives a contribution to the operator
\begin{equation}
  {\cal L} = \frac{\tilde z}{m_\eta^2} \bar u_R^\alpha \gamma^\mu c_R^\alpha \, \bar u_R^\beta \gamma_\mu c_R^\beta
\end{equation}
given by
\begin{equation}
 \tilde z = - \frac{f_1^2f_2^2}{96\pi^2} g_\chi(m_{\chi}^2/m_\eta^2)
\end{equation}
where $g_\chi(x)=24xf_6(x)+12\tilde f_6(x)$  (with $g_\chi(1)=4/5$). The functions $f_6(x)$ and $\tilde f_6(x)$ are given e.g. in \cite{Gedalia:2009kh}.
On the other hand, the experimental constraint inferred from measurements of $\Delta m_D$ is $|\tilde z|\lesssim 5.7\cdot 10^{-7}(m_\eta/\TeV)^2$ \cite{Gedalia:2009kh}. For $m_\eta\simeq m_{\chi}$, this translates into an upper bound
\begin{equation}
  |f_2/f_1| \lesssim 0.026 \times (f_1)^{-2} \times \frac{m_\eta}{\TeV} \;.
\end{equation}
Since thermal production requires typically $f_1\sim {\cal O}(1)$, this means that $\eta$ has to couple nearly exclusively to the up-quark, with very suppressed coupling to charm (or vice-versa). A possible exception are regions in parameter space with strong coannihilation for which $f_1 \ll 1$. For a generic $f_1\sim {\cal O}(1)$, the flavour-vector $f_i$ should be aligned with the mass eigenbasis of the quarks. This can be realized e.g. in the presence of a $U(1)$ flavour symmetry under which $u_{R,L}$ and $\eta$ transform with equal charge, while all other states are uncharged. This symmetry is then broken only by the CKM mixing in the left-handed quark sector, and thus this breaking should lead to a misalignment suppressed by the quark masses as well as CKM mixing angles. More precisely, one may consider a situation where $f_i\propto (1,0,0)$ at some high scale $M$. Due to renormalization group running, the quark mass matrices $M_u(\mu)$ and $M_d(\mu)$ are scale-dependent. This leads to a running of the diagonalization matrices $M_u^{diag}(\mu)=V_u^L(\mu)M_u(\mu)V_u^R(\mu)^\dag$, with a similar expression for the down-type quarks. The left-handed rotations lead to the well-known running of the CKM matrix $V_{CKM}(\mu)=V_u^L(\mu)V_d^L(\mu)^\dag$, while the right-handed rotations are unobservable in the Standard Model \cite{Ma:1979cw}. However, in the present case they lead to a flavour-dependent running of the dark matter couplings,
\begin{equation}
 f_i(\mu)=V_u^R(\mu)_{ij} f_j(M) \,,
\end{equation}
where we neglect flavour-insensitive contributions to the running and assume that $V_u^R(M)_{ij}=\delta_{ij}$. Using the one-loop RGEs for the quark mass matrices from \cite{Barger:1992pk}, one finds for the off-diagonal entry corresponding to $i=u$ and $j=c$
\begin{equation}
\frac{d}{d\ln \mu} V_u^R(\mu)^\dag_{uc} = - \frac{3}{ 16\pi^2 v_{EW}^2 } \frac{m_um_c}{m_u^2-m_c^2} \left(V_{ud}V_{cd}^* m_d^2 + V_{us}V_{cs}^* m_s^2 + V_{ub}V_{cb}^* m_b^2 \right)
\end{equation}
where $V_{ud}$ etc. denotes the CKM matrix elements.
Thus, even for perfect alignment $f_i(M)=(f_1,0,0)$ at the high scale, the coupling to the second generation induced by the running is approximately
\begin{equation}
  |f_2/f_1| \simeq |V_u^R(\mu)_{uc}| \sim \frac{3}{16\pi^2}\frac{m_u}{m_c}\frac{|V_{us}V_{cs}^* m_s^2 + V_{ub}V_{cb}^* m_b^2|}{v_{EW}^2} \ln\frac{M}{\mu} \sim 10^{-10}
\end{equation}
which is safely below the upper bound required from $D-\bar D$ mixing.

Alternatively, one may consider a situation where three additional scalars $\eta_i$ are introduced, which are taken to transform under the
$SU(3)_{u_R}$ flavour symmetry. Then the allowed coupling is of the form
\begin{equation}
  {\cal L} = -f \sum_i \bar u_{Ri} \chi \eta_i
\end{equation}
and the $\eta_i$ are all mass-degenerate. One may consider a breaking of the symmetry in the scalar mass matrix, which induces non-degenerate mass eigenvalues of the $\eta_i$, and singles out a preferred basis, namely the mass eigenbasis (similar to the right-handed squarks in the MSSM). After rotating into this basis (as well as the mass basis for the quarks) the interaction term has the generic form
\begin{equation}
  {\cal L} = -f K_{ij} \sum_i \bar u_{Ri} \chi \eta_j
\end{equation}
where $K$ is a unitary matrix, which can have large off-diagonal entries. The resulting contribution to the box diagram will be proportional to \cite{Nir:1993mx}
\begin{equation}
  \sum_{\alpha,\beta} K_{1\alpha}K_{2\alpha}^* K_{1\beta}K_{2\beta}^* F(m_{\eta_\alpha},m_{\eta_\beta})
\end{equation}
where $F$ is a function of the masses. In the limit of degenerate masses this expression goes to zero by virtue of the unitarity condition $(KK^\dag)_{12}=0$. Lets assume for concreteness that the mixing with the third generation is negligible, similar as in the CKM matrix.
In this case the box diagram gives a contribution 
\begin{equation}
  \tilde z = - \frac{f^4}{384\pi^2} g_\chi(m_\chi^2/m_\eta^2) \times \delta^2
\end{equation}
with $\delta=K_{21}K_{11}(m_{\eta_1}^2-m_{\eta_2}^2)/m_{\eta}^2$ and $m_\eta=(m_{\eta_1}+m_{\eta_2})/2$. Thus, the strong requirement of precise alignment $f_2/f_1=K_{21}/K_{11}$ found above can be considerably relaxed provided the masses are quasi-degenerate. For order one mixing, the upper bound on $\tilde z$ required from $D-\bar D$ mixing then translates into an upper bound on the mass splitting
\begin{equation}
  \frac{|m_{\eta_1}-m_{\eta_2}|}{m_{\eta_1}+m_{\eta_2}} \lesssim 0.026 \times (f_1)^{-2} \times \frac{m_\eta}{\TeV} \;.
\end{equation}
Thus, in both cases discussed above, the flavour constraints can be fulfilled in presence of an (approximate) flavour symmetry.

\end{appendix}

\providecommand{\href}[2]{#2}\begingroup\raggedright\endgroup

\end{document}